\documentclass[12pt]{article}

\usepackage{graphicx,amssymb,amsmath,dsfont,txfonts} 
\usepackage{bbm} 
\usepackage{mathtools} 

\usepackage[a4paper]{geometry} 
\def\dfrac#1#2{\frac{\displaystyle\strut #1}{\displaystyle\strut #2}}
\DeclareMathOperator{\tr}{tr}
\def\bra#1{\mathinner{\langle{#1}|}}
\def\ket#1{\mathinner{|{#1}\rangle}}
\def\braket#1{\mathinner{\langle{#1}\rangle}}
\renewcommand\Re{\operatorname{\mathfrak{R\hspace{-0.1em}e}}}
\newcommand{\rchi}{\raisebox{0.19em}{$\chi$}}

\begin{document}

\title{\LARGE\bf 
Quantum parameter estimation \\ on coherently superposed noisy channels}

\author{Fran\c{c}ois {\sc Chapeau-Blondeau}, \\
    Laboratoire Angevin de Recherche en Ing\'enierie des Syst\`emes (LARIS), \\
    Universit\'e d'Angers,
    62 avenue Notre Dame du Lac, 49000 Angers, France.
}

\date{\today}

\maketitle

\parindent=8mm \parskip=0ex

\begin{abstract}
A generic qubit unitary operator affected by quantum noise is duplicated and inserted in a 
coherently superposed channel, superposing two paths offered to a probe qubit across the noisy 
unitary, and driven by a control qubit. A characterization is performed of the transformation 
realized by the superposed channel on the joint state of the probe-control qubit pair. The 
superposed channel is then specifically analyzed for the fundamental metrological task of 
phase estimation on the noisy unitary, with the performance assessed by the Fisher information, 
classical or quantum. A comparison is made with conventional estimation techniques and also with 
a quantum switched channel with indefinite causal order recently investigated for a similar task 
of phase estimation. In the analysis here, a first important observation is that the control qubit 
of the superposed channel, although it never directly interacts with the unitary being estimated, 
can nevertheless be measured alone for effective estimation, while discarding the probe qubit 
that interacts with the unitary. This property is also present with the switched channel but is 
inaccessible with conventional techniques. The optimal measurement of the control qubit here
is characterized in general conditions. A second important observation is that the noise 
plays an essential role in coupling the control qubit to the unitary, and that the control qubit 
remains operative for phase estimation at very strong noise, even with a fully depolarizing 
noise, whereas conventional estimation and the switched channel become inoperative in these 
conditions. The results extend the analysis of the capabilities of coherently controlled 
channels which represent novel devices exploitable for quantum signal and information 
processing.
\end{abstract}

\maketitle

\section{Introduction}

{\let\thefootnote\relax\footnote{{Preprint of a paper published by {\em Physical Review A},
vol.~104, 032214, pp.~1--16 (2021). \\
https://doi.org/10.1103/PhysRevA.104.032214 \hfill 
https://journals.aps.org/pra/abstract/10.1103/PhysRevA.104.032214}}}
Quantum channels, when exploited for information processing, can be combined in specifically
quantum ways, that differ from standard classical combinations, and that offer novel
capabilities with no classical analogue. Such techniques have been recently introduced,
and two basic and prominent examples of them are the switch of two quantum channels in
indefinite causal order, and the coherent superposition of two quantum channels. These two 
basic schemes combine two quantum channels (1) and (2) and operate in the following manner.

The first scheme (the switched channel) \cite{Oreshkov12,Chiribella13,Ebler18} uses a quantum 
switch driven by a control qubit to select between cascading the two channels in the order 
(1)--(2) or (2)--(1). When the control qubit is placed in a superposed state, the two individual 
channels become cascaded in a superposition of the two classical orders (1)--(2) and (2)--(1), 
realizing a switched quantum channel incorporating simultaneously the two alternative orders, or 
with indefinite causal order. 

The second scheme (the superposed channel) \cite{Abbott20,Chiribella19,Kristjansson20}
uses a control qubit to drive an information-carrying 
or signal quantum state either across channel (1) or across channel (2). When the control qubit 
is placed in a superposed state, the scheme realizes a coherently superposed channel offering a 
coherent superposition of the two alternative paths across (1) and (2).

Both schemes are specifically quantum devices, and represent novel resources exploitable for 
quantum information processing \cite{Chiribella19,Kristjansson20}. The main area where these 
novel techniques have been analyzed and compared is quantum communication. A typical and striking 
benefit is when the two channels (1) and (2) are two completely depolarizing quantum channels, 
which are therefore individually incapable of transmitting any useful information. It has then 
been shown that when two such channels are associated into a switched channel with indefinite 
order \cite{Ebler18,Loizeau20}, or into a coherently superposed channel 
\cite{Abbott20,Loizeau20}, in each scheme effective information communication becomes possible.
Comparison has been performed and discussed \cite{Abbott20,Loizeau20,Chiribella21} to better 
appreciate the mechanisms and specificities of the two schemes, and the respective roles of 
indefinite order and of coherent superposition, particularly in reaching similar capabilities 
of information communication through depolarizing channels.

More recently, switched channels with indefinite order have been investigated in another 
important information-processing task, for quantum metrology, consisting in phase estimation
from a noisy unitary transformation \cite{Chapeau21}. Specific capabilities, useful to 
estimation, have been reported and analyzed in the switched channels, for example the 
possibility of using a completely depolarized input signal as an operative probe for estimation 
(while this is never possible with conventional techniques).

As a complement to \cite{Chapeau21}, we will investigate a comparable scenario of phase 
estimation from a noisy unitary, when undertaken here by means of a coherently superposed channel.
The study will bring additional results on the capabilities of coherently superposed channels
for quantum information processing, along with more elements of comparison between the two
schemes of switched indefinite order and of coherently superposed channels.

\section{Coherent superposition of two quantum channels} \label{switch_sec}

\begin{figure}[htb]
\centerline{\includegraphics[width=70mm]{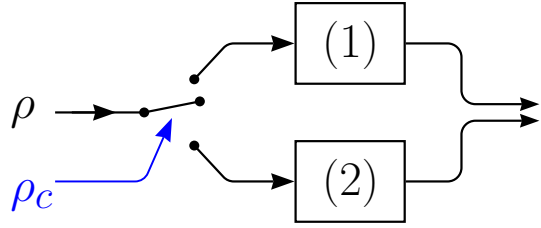}}
\caption[what appears in lof LL p177]
{A quantum state $\rho$ as an input signal can be sent across either channel (1) or channel (2), 
according to the state $\rho_c$ of a control qubit.
}
\label{figSwi1}
\end{figure}

We consider as in \cite{Abbott20,Chiribella19,Kristjansson20}, acting on quantum systems with 
Hilbert space $\mathcal{H}$, a quantum channel (1) with $K_1$ Kraus operators 
$\mathsf{K}_j^{(1)} \in \mathcal{L}(\mathcal{H})$ for $j=1$ to $K_1$, and a second quantum 
channel (2) with $K_2$ Kraus operators $\mathsf{K}_k^{(2)} \in \mathcal{L}(\mathcal{H})$ for 
$k=1$ to $K_2$. A quantum system with density operator $\rho \in \mathcal{L}(\mathcal{H})$ 
acting as an input signal can be sent across either channel (1) or channel (2), according to 
the state of a control qubit with two-dimensional Hilbert space $\mathcal{H}_2$ referred to 
the orthonormal basis $\bigl\{\ket{0_c}, \ket{1_c} \bigr\}$. When the control qubit is in state 
$\ket{0_c}$ the input signal $\rho$ is sent across channel (1), and when the control qubit is 
in state $\ket{1_c}$ the input signal $\rho$ is sent across channel (2). The resulting quantum 
superposed channel is depicted in Fig.~\ref{figSwi1}.

In this study, the coherently superposed channel of Fig.~\ref{figSwi1} will superpose two noisy 
unitary channels to be involved in a task of quantum parameter estimation, the same task 
considered in \cite{Chapeau21} in a switched channel with indefinite causal order. Before we 
come to this, the operation of the coherently superposed channel of Fig.~\ref{figSwi1} needs to 
be more completely defined, especially for an arbitrary state $\rho_c$ of the control qubit. For 
this purpose, following \cite{Abbott20}, we will rely on a Stinespring dilation for the 
superposed channel of Fig.~\ref{figSwi1}, involving a description of the interaction with the 
surrounding environment underlying the operation of the superposed channel. 

The compound signal-control-environment, with a control qubit in state $\ket{0_c}$ evolves 
unitarily as
\begin{equation}
\ket{\psi}\otimes \ket{0_c}\otimes \ket{g^{(1)}} \otimes \ket{g^{(2)}} \longmapsto 
\sum_{j=1}^{K_1} \mathsf{K}_j^{(1)} \ket{\psi} \otimes \ket{0_c}
\otimes \ket{e_j^{(1)}} \otimes \ket{g^{(2)}} \;,
\label{environ1}
\end{equation}
and with a control qubit in state $\ket{1_c}$ it evolves unitarily as
\begin{equation}
\ket{\psi}\otimes \ket{1_c}\otimes \ket{g^{(1)}} \otimes \ket{g^{(2)}} \longmapsto 
\sum_{k=1}^{K_2} \mathsf{K}_k^{(2)} \ket{\psi} \otimes \ket{1_c}
\otimes \ket{g^{(1)}} \otimes \ket{e_k^{(2)}} \;,
\label{environ2}
\end{equation}
for any signal state $\ket{\psi}\in \mathcal{H}$, and where we have introduced for channel (1) 
with $K_1$ Kraus operators a $K_1$-dimensional environment $E_1$ of orthonormal basis 
$\bigl\{\ket{e_j^{(1)}} \bigr\}_{j=1}^{K_1}$ initialized in state $\ket{g^{(1)}}$, and for 
channel (2) with $K_2$ Kraus operators a $K_2$-dimensional environment $E_2$ of orthonormal 
basis $\bigl\{\ket{e_k^{(2)}} \bigr\}_{k=1}^{K_2}$ initialized in state $\ket{g^{(2)}}$.
In practice, a controlled operation like in Eqs.~(\ref{environ1})--(\ref{environ2}), can
be carried out by an interferometric setup, such as those discussed in 
\cite{Zhou11,Friis14,Abbott20}, with for instance the control realized by the polarization of a 
photon and the signal state $\ket{\psi}$ by other degrees of freedom such as quantum spatial
modes of the same photon.

An arbitrary state $\rho_c$ of the control qubit is defined as a linear combination of the four 
matrix elements $\ket{0_c}\bra{0_c}$, $\ket{0_c}\bra{1_c}$, $\ket{1_c}\bra{0_c}$ and 
$\ket{1_c}\bra{1_c}$, and in such circumstance the corresponding evolution of the compound
signal-control-environment follows by linearity of Eqs.~(\ref{environ1})--(\ref{environ2}). For 
instance, for the control $\ket{0_c}\bra{0_c}$ we have from Eq.~(\ref{environ1}),
\begin{equation}
\ket{\psi}\bra{\psi}\otimes \ket{0_c}\bra{0_c}\otimes \ket{g^{(1)}}\bra{g^{(1)}} \otimes 
\ket{g^{(2)}}\bra{g^{(2)}} \longmapsto 
\sum_{j=1}^{K_1}\sum_{j'=1}^{K_1}\mathsf{K}_j^{(1)}\ket{\psi}\bra{\psi} \mathsf{K}_{j'}^{(1)\dagger} 
\otimes \ket{0_c}\bra{0_c} \otimes \ket{e_j^{(1)}}\bra{e_{j'}^{(1)}} \otimes 
\ket{g^{(2)}}\bra{g^{(2)}} \;,
\label{environ3}
\end{equation}
which upon partial tracing over the two environments via $\tr_{E_1 E_2}(\cdot)$ leads to the 
non-unitary evolution of the signal-control compound as
\begin{equation}
\ket{\psi}\bra{\psi} \otimes \ket{0_c}\bra{0_c}  \longmapsto 
\sum_{j=1}^{K_1} \mathsf{K}_j^{(1)} \ket{\psi}\bra{\psi} \mathsf{K}_{j}^{(1)\dagger} 
\otimes \ket{0_c}\bra{0_c}  \;,
\label{environ4}
\end{equation}
where we recognize for the signal state $\ket{\psi}\bra{\psi} $ the evolution induced by channel 
(1) alone. In a similar way, the evolution with the control $\ket{1_c}\bra{1_c}$ and 
Eq.~(\ref{environ2}) leads to
\begin{equation}
\ket{\psi}\bra{\psi} \otimes \ket{1_c}\bra{1_c}  \longmapsto 
\sum_{k=1}^{K_2} \mathsf{K}_k^{(2)} \ket{\psi}\bra{\psi} \mathsf{K}_{k}^{(2)\dagger} 
\otimes \ket{1_c}\bra{1_c}  \;,
\label{environ4b}
\end{equation}
where we recognize for the signal state $\ket{\psi}\bra{\psi}$ the evolution induced by channel 
(2) alone.

With the control $\ket{0_c}\bra{1_c}$ we have from Eqs.~(\ref{environ1}) and 
(\ref{environ2}),
\begin{equation}
\ket{\psi}\bra{\psi}\otimes \ket{0_c}\bra{1_c}\otimes \ket{g^{(1)}}\bra{g^{(1)}} \otimes 
\ket{g^{(2)}}\bra{g^{(2)}} \longmapsto 
\sum_{j=1}^{K_1}\sum_{k=1}^{K_2} \mathsf{K}_j^{(1)}\ket{\psi}\bra{\psi} \mathsf{K}_{k}^{(2)\dagger}
\otimes \ket{0_c}\bra{1_c} \otimes \ket{e_j^{(1)}}\bra{e_{0}^{(1)}} \otimes 
\ket{g^{(2)}}\bra{e_k^{(2)}} \;,
\label{environ5}
\end{equation}
which by partial tracing over the two environments $E_1E_2$ leads for the signal-control compound 
to the transformed state
\begin{equation}
\left( \sum_{j=1}^{K_1} \braket{g^{(1)} | e_j^{(1)}} \mathsf{K}_j^{(1)} \right)
\ket{\psi}\bra{\psi} 
\left( \sum_{k=1}^{K_2} \braket{e_k^{(2)} | g^{(2)}} \mathsf{K}_k^{(2)^\dagger} \right)
\otimes \ket{0_c}\bra{1_c} =
\mathsf{T}_1 \ket{\psi}\bra{\psi} \mathsf{T}_2^\dagger \otimes \ket{0_c}\bra{1_c} \;,
\label{environ6}
\end{equation}
where we have defined the two operators of $\mathcal{L}(\mathcal{H})$,
\begin{eqnarray}
\label{opT1}
\mathsf{T}_1 &=&\sum_{j=1}^{K_1} \braket{g^{(1)} | e_j^{(1)}} \mathsf{K}_j^{(1)} \;,\\
\label{opT2}
\mathsf{T}_2 &=&\sum_{k=1}^{K_2} \braket{g^{(2)} | e_k^{(2)}} \mathsf{K}_k^{(2)} \;,
\end{eqnarray}
that are the two transformation operators introduced in \cite{Abbott20}.
The operators $\mathsf{T}_1$ and $\mathsf{T}_2$ of Eqs.~(\ref{opT1})--(\ref{opT2}) appear as 
essential to the operation of the superposed channel of Fig.~\ref{figSwi1}. The two operators 
$\mathsf{T}_1$ and $\mathsf{T}_2$, as visible in Eqs.~(\ref{opT1})--(\ref{opT2}), depend on 
the environment models for implementing channels (1) and (2), especially via the initial states 
$\ket{g^{(1)}}$ and $\ket{g^{(2)}}$. This is the same observation as in \cite{Abbott20}, that 
the Kraus operators $\{\mathsf{K}_j^{(1)} \}_{j=1}^{K_1}$ and $\{\mathsf{K}_k^{(2)} \}_{k=1}^{K_2}$ 
alone, (unlike the situation of the switched channel of 
\cite{Oreshkov12,Chiribella13,Ebler18,Chapeau21}), are not sufficient to precisely specify the 
operation of the superposed channel of Fig.~\ref{figSwi1}, but that a reference to a specific 
implementation of each channel via an environment model is necessary. We will discuss further 
these specific aspects related to the underlying channel implementation and relevant to the 
superposed channel, in particular in Section~\ref{implem_sec} and in the Appendix. Now we 
proceed in the analysis of the coherently superposed channel.

For an arbitrary input signal state $\rho \in \mathcal{L}(\mathcal{H})$, which is a convex sum 
of pure states like $\ket{\psi}\bra{\psi}$, and an arbitrary state 
$\rho_c \in \mathcal{L}(\mathcal{H}_2)$ for the control qubit, we consequently obtain by 
linearity of the evolution, the quantum operation realized on the signal-control compound by 
the coherently superposed channel of Fig.~\ref{figSwi1}, as
\begin{eqnarray}
\nonumber
\mathcal{S}(\rho \otimes \rho_c) &=& 
\mathcal{S}_{00}(\rho) \otimes \braket{0_c|\rho_c|0_c}\ket{0_c}\bra{0_c} +
\mathcal{S}_{01}(\rho) \otimes \braket{0_c|\rho_c|1_c}\ket{0_c}\bra{1_c} \\
\label{Sgen2} 
\mbox{} &+& \mathcal{S}_{01}^\dagger(\rho) \otimes \braket{1_c|\rho_c|0_c}\ket{1_c}\bra{0_c} +
\mathcal{S}_{11}(\rho) \otimes \braket{1_c|\rho_c|1_c}\ket{1_c}\bra{1_c} \;,
\end{eqnarray}
with the superoperators
\begin{eqnarray}
\label{S00}
\mathcal{S}_{00}(\rho) &=& \sum_{j=1}^{K_1} \mathsf{K}_j^{(1)} \rho 
\mathsf{K}_j^{(1)\dagger}  \;,\\
\label{S11}
\mathcal{S}_{11}(\rho) &=& \sum_{k=1}^{K_2} \mathsf{K}_k^{(2)} \rho 
\mathsf{K}_k^{(2)\dagger}  \;,\\
\label{S01}
\mathcal{S}_{01}(\rho) &=& \mathsf{T}_1 \rho \mathsf{T}_2^\dagger \;.
\end{eqnarray}

The superoperator $\mathcal{S}_{00}(\rho)$ describes the quantum operation realized by channel 
(1) alone, and similarly with $\mathcal{S}_{11}(\rho)$ for channel (2). The superoperator 
$\mathcal{S}_{01}(\rho)$ is an interference term manifesting the coherent superposition of 
channels (1) and (2) in Fig.~\ref{figSwi1}. In the joint state $\mathcal{S}(\rho \otimes \rho_c)$ 
of Eq.~(\ref{Sgen2}), if the control qubit were discarded (unobserved) and traced out, the 
resulting quantum operation on $\rho$ would reduce to $\rho \mapsto
\mathcal{S}_{00}(\rho) \braket{0_c|\rho_c|0_c}+ \mathcal{S}_{11}(\rho)\braket{1_c|\rho_c|1_c}$,
representing a classical probabilistic (convex) combination of the channels (1) and (2) traversed 
with respective probabilities $\braket{0_c|\rho_c|0_c}$ and $\braket{1_c|\rho_c|1_c}$.
By contrast, if the control qubit is treated coherently with $\rho$, it can give rise to specific, 
specifically quantum, behaviors from the superposed channel, as we are going to see.

An interesting and specifically quantum feature is that the control qubit can be placed in the 
superposed state $\ket{\psi_c}=\sqrt{p_c}\ket{0_c}+\sqrt{1-p_c}\ket{1_c}$, with $p_c \in [0, 1]$.
This produces in Fig.~\ref{figSwi1} a quantum superposed channel representing a coherent 
superposition of two alternative paths for $\rho$ across channel (1) and channel (2) 
simultaneously. With $\rho_c=\ket{\psi_c}\bra{\psi_c}$, the quantum operation resulting from 
Eq.~(\ref{Sgen2}) takes the form
\begin{eqnarray}
\nonumber
\mathcal{S}(\rho \otimes \rho_c) &=& p_c
\mathcal{S}_{00}(\rho) \otimes \ket{0_c}\bra{0_c} + (1-p_c)
\mathcal{S}_{11}(\rho) \otimes \ket{1_c}\bra{1_c} \\
\label{Sgenc}
&+& \sqrt{(1-p_c)p_c} \,\Bigl[ \mathcal{S}_{01}(\rho) \otimes \ket{0_c}\bra{1_c} + 
\mathcal{S}_{01}^\dagger(\rho) \otimes \ket{1_c}\bra{0_c} \Bigr] \;.
\end{eqnarray}

This description of the operation of the coherently superposed channel of Fig.~\ref{figSwi1} 
applies when (1) and (2) are two arbitrary quantum channels.
We will now consider the situation as in \cite{Chapeau21}, where the quantum channels (1) and (2) 
are qubit channels, under a form which is often encountered in quantum metrology, and consisting 
in a unitary operator $\mathsf{U}_\xi$ affected by a quantum noise $\mathcal{N}(\cdot)$.

\section{A qubit unitary channel with noise} \label{unqbit_sec}

When channels (1) and (2) in Fig.~\ref{figSwi1} are qubit channels, an interesting feature is 
that the two transformation operators $\mathsf{T}_1$ and $\mathsf{T}_2$ of Eqs.~(\ref{opT1}) 
and (\ref{opT2}), are operators of $\mathcal{L}(\mathcal{H}_2)$ which can be conveniently 
characterized in Bloch representation, i.e.\ in the basis of the four Pauli operators 
$\{\mathrm{I}_2, \sigma_x, \sigma_y, \sigma_z \}$ forming an orthogonal basis for 
the operator space $\mathcal{L}(\mathcal{H}_2)$. The Bloch representation for $\mathsf{T}_1$ and 
$\mathsf{T}_2$ will originate in their respective set of Kraus operators and underlying
environment models appearing in Eqs.~(\ref{opT1}) and (\ref{opT2}), but will generally involve 
no more than four coordinates, for each operator $\mathsf{T}_1$ or $\mathsf{T}_2$, in the Pauli 
basis. This will usually enable a concise characterization in Bloch representation of the 
superoperator $\mathcal{S}_{01}(\rho)$ of Eq.~(\ref{S01}) that conveys the coupling and effects 
specific to the coherently superposed channel. This approach, in its generality for qubit 
channels, will be presented here in the interesting case where each channel (1) and (2) is a 
unitary qubit channel affected by a generic qubit noise.

\bigbreak
For qubits with two-dimensional Hilbert space $\mathcal{H}_2$, the density operator is 
represented in Bloch representation \cite{Nielsen00} under the form
\begin{equation}
\rho =\frac{1}{2}\bigl( \mathrm{I}_2 + \vec{r}\cdot \vec{\sigma} \bigr) \;,
\label{roBloch}
\end{equation}
where $\mathrm{I}_2$ is the identity operator on $\mathcal{H}_2$, and $\vec{\sigma}$ a formal 
vector assembling the three (traceless Hermitian unitary) Pauli operators 
$[\sigma_x, \sigma_y, \sigma_z]=\vec{\sigma}$. The Bloch vector $\vec{r} \in \mathbbm{R}^3$ 
characterizing the density operator has norm $\| \vec{r}\, \|=1$ for a pure state, and 
$\| \vec{r}\, \|<1$ for a mixed state.

A qubit unitary operator $\mathsf{U}_\xi$ is introduced with the general parameterization
\cite{Nielsen00} 
\begin{equation}
\mathsf{U}_\xi=\exp\Bigl(-i\dfrac{\xi}{2} \vec{n} \cdot \vec{\sigma} \Bigr) \;,
\label{Uxi}
\end{equation}
where $\vec{n}=[n_x, n_y, n_z]^\top$ is a unit vector of $\mathbbm{R}^3$, and $\xi$ a phase 
angle in $[0, 2\pi )$. Such an $\mathsf{U}_\xi$ represents for instance the transformation of a 
photon polarization by an optical interferometer, with the axis $\vec{n}$ fixed by the orientation 
of the polarizing beam splitter and a phase shift $\xi$ occurring between the two arms of the 
interferometer. From a qubit state $\rho$ in Bloch representation as in Eq.~(\ref{roBloch}), the 
unitary $\mathsf{U}_\xi$ produces the transformed state
\begin{equation}
\mathsf{U}_\xi \rho \mathsf{U}_\xi^\dagger = 
\frac{1}{2}\bigl( \mathrm{I}_2 + U_\xi\vec{r}\cdot \vec{\sigma} \bigr) \;,
\label{Uxi_ro}
\end{equation}
where $U_\xi$ is the $3\times 3$ real matrix\footnote{We use the notation $\mathsf{U}_\xi$ in 
upright font for the unitary operator acting in the complex Hilbert space $\mathcal{H}_2$ of the 
qubit; while we use the notation $U_\xi$ in italic font for the real matrix expressing the action 
of the unitary operator in the Bloch representation of qubit states in $\mathbbm{R}^3$.}
representing in $\mathbbm{R}^3$ the rotation around the axis $\vec{n}$ by the angle $\xi$.

\bigbreak
A qubit noise $\mathcal{N}(\cdot)$ is also introduced, implementing a quantum operation which
is represented \cite{Nielsen00} by the Kraus operators $\Lambda_j$ of 
$\mathcal{L}(\mathcal{H}_2)$, which need not be more than four for representing any qubit noise.
Equivalently, the transformation $\rho \mapsto \mathcal{N}(\rho)$ by the qubit noise can be 
defined \cite{Nielsen00} by the affine transformation of the qubit Bloch vector
\begin{equation}
\vec{r} \longmapsto A\vec{r} + \vec{c} \;,
\label{Pauli1}
\end{equation}
mapping the unit Bloch ball into itself, and characterized by the $3\times 3$ real matrix
$A$ and vector $\vec{c} \in \mathbbm{R}^3$.

\begin{figure}[htb]
\centerline{\includegraphics[width=70mm]{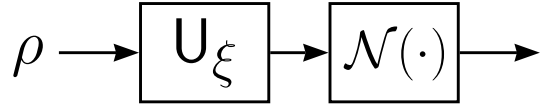}}
\caption[what appears in lof LL p177]
{A qubit channel formed by the unitary transformation $\mathsf{U}_\xi$ of Eq.~(\ref{Uxi}) and 
the general qubit noise $\mathcal{N}(\cdot)$ of Eq.~(\ref{Pauli1}). As a whole 
this channel is an instance of channel (1) or (2) considered in Fig.~\ref{figSwi1}.
}
\label{figUxiN}
\end{figure}

Each quantum channel like (1) or (2) of Section~\ref{switch_sec} is formed by cascading the 
unitary transformation $\mathsf{U}_\xi$ of Eq.~(\ref{Uxi}) and the general qubit noise 
$\mathcal{N}(\cdot)$ of Eq.~(\ref{Pauli1}), as depicted in Fig.~\ref{figUxiN}.

For the quantum channel of Fig.~\ref{figUxiN}, the Kraus operators like 
$\mathsf{K}_j^{(1)}$ or $\mathsf{K}_k^{(2)}$ of Section~\ref{switch_sec} result as
$\mathsf{K}_j = \Lambda_j \mathsf{U}_\xi$. Equivalently, for a qubit state $\rho$ in Bloch 
representation as in Eq.~(\ref{roBloch}), the cascade of $\mathsf{U}_\xi$ then 
$\mathcal{N}(\cdot)$ produces the transformed state
\begin{equation}
\mathcal{N}(\mathsf{U}_\xi \rho \mathsf{U}_\xi^\dagger) = 
\frac{1}{2}\bigl[ \mathrm{I}_2 + ( A U_\xi\vec{r}+\vec{c}\,) \cdot \vec{\sigma} \bigr] \;.
\label{NUxi_ro}
\end{equation}

\section{Coherent superposition of two noisy unitaries} \label{qbswitch_sec}

Two identical qubit channels formed by $\mathsf{U}_\xi$ and $\mathcal{N}(\cdot)$ as in 
Fig.~\ref{figUxiN} are associated as in Fig.~\ref{figSwi1} through the coherent quantum 
superposition of Section~\ref{switch_sec}, with however two independent sources of a same 
type of qubit noise $\mathcal{N}(\cdot)$ according to Eq.~(\ref{Pauli1}). For two such identical 
channels (1) and (2), one has $\mathcal{S}_{00}(\rho)=\mathcal{S}_{11}(\rho)$ in 
Eqs.~(\ref{S00})--(\ref{S11}), and also $\mathcal{S}_{01}^\dagger(\rho)=\mathcal{S}_{01}(\rho)$
in Eq.~(\ref{S01}). On the probe qubit in state $\rho$ and control qubit in state
$\rho_c=\ket{\psi_c}\bra{\psi_c}$, the coherently superposed channel therefore realizes 
the two-qubit quantum operation from Eq.~(\ref{Sgenc}) reading
\begin{eqnarray}
\nonumber
\mathcal{S}(\rho \otimes \rho_c) &=& \mathcal{S}_{00}(\rho) \otimes \bigl[ 
p_c \ket{0_c}\bra{0_c} + (1-p_c) \ket{1_c}\bra{1_c} \bigr]\\
\label{Sgenqb}
&+& \mathcal{S}_{01}(\rho) \otimes \sqrt{(1-p_c)p_c} \bigl( \ket{0_c}\bra{1_c} + 
\ket{1_c}\bra{0_c} \bigr) \;.
\end{eqnarray}

As already mentioned, the superoperator $\mathcal{S}_{00}(\rho)$ (and $\mathcal{S}_{11}(\rho)$
similarly) describes the quantum operation realized on $\rho$ by directly traversing the noisy 
unitary channel of Fig.~\ref{figUxiN}, and it is given in Bloch representation by 
Eq.~(\ref{NUxi_ro}) as
\begin{equation}
\mathcal{S}_{00}(\rho)= 
\frac{1}{2}\bigl[ \mathrm{I}_2 + ( A U_\xi\vec{r}+\vec{c}\,) \cdot \vec{\sigma} \bigr] \;.
\label{NUNU_ro}
\end{equation}

In the superoperator $\mathcal{S}_{01}(\rho)$ of Eq.~(\ref{S01}), for two identical channels (1) 
and (2) with the same set of Kraus operators $\mathsf{K}_j = \Lambda_j \mathsf{U}_\xi$, the two 
transformation operators $\mathsf{T}_1$ and $\mathsf{T}_2$ of Eqs.~(\ref{opT1})--(\ref{opT2}) 
have the same form, and they can be expressed as 
$\mathsf{T}_1 = \mathsf{T}_2 = \mathsf{T} \mathsf{U}_\xi$, with the operator of 
$\mathcal{L}(\mathcal{H}_2)$,
\begin{equation}
\mathsf{T} = \sum_{j} \braket{g | e_j} \Lambda_j \;,
\label{opTT_1}
\end{equation}
where $\bigl\{ \ket{e_j} \bigr\}$ is the reference basis for each of the two identical 
environment models initialized in state $\ket{g}$.
As we announced at the beginning of Section~\ref{unqbit_sec}, it is generally possible to
represent the operator $\mathsf{T} \in \mathcal{L}(\mathcal{H}_2)$ in the Pauli basis with
the Bloch representation
\begin{equation}
\mathsf{T} = t_0\mathrm{I}_2 + \vec{t}\cdot \vec{\sigma} \;,
\label{opTT}
\end{equation}
with the four complex coordinates $t_0$ and $\vec{t}=[t_x, t_y, t_z]^\top$. For example, for the 
class of Pauli noises \cite{Nielsen00,Wilde17} the four Kraus operators $\{\Lambda_j\}$ are
$\bigl\{\sqrt{p_0}\mathrm{I}_2, \sqrt{p_x}\sigma_x, \sqrt{p_y}\sigma_y, \sqrt{p_z}\sigma_z \bigr\}$
with $\{p_j \}$ a probability distribution. This class of Pauli noises acts through random 
applications of the four Pauli operators, and it contains in particular such important noises as 
the bit-flip, the phase-flip, the bit-phase-flip, the depolarizing noises. 
The corresponding Bloch coordinates for $\mathsf{T}$ in Eq.~(\ref{opTT}) are
$t_0 = \braket{g | e_1}\!\sqrt{p_0}$ and $\vec{t}=\bigl[\braket{g | e_2}\!\sqrt{p_x}, 
\braket{g | e_3}\!\sqrt{p_y}, \braket{g | e_4}\!\sqrt{p_z} \,\bigr]^\top$.

It is now possible to characterize the action of the superoperator $\mathcal{S}_{01}(\rho)$ of 
Eq.~(\ref{S01}) in Bloch representation, starting with
\begin{equation}
\mathcal{S}_{01}(\rho) = \frac{1}{2} \mathcal{S}_{01}(\mathrm{I}_2) +
\frac{1}{2} \mathcal{S}_{01}(\vec{r}\cdot \vec{\sigma}  \,) =
\mathsf{T} \mathsf{U}_\xi \rho \mathsf{U}_\xi^\dagger \mathsf{T}^\dagger \;.
\label{S01_T}
\end{equation}
To proceed, we will need the identity 
$(\vec{a}\cdot\vec{\sigma}\,)(\vec{b}\cdot\vec{\sigma}\,)=(\vec{a}\,\vec{b}\,)\,\mathrm{I}_2 +
i\,(\vec{a}\times \vec{b}\,) \cdot \vec{\sigma}$ for any two vectors $\vec{a}$ and $\vec{b}$ of 
$\mathbbm{C}^3$.
As a next step, we can evaluate $\mathcal{S}_{01}(\mathrm{I}_2)=\mathsf{T}\mathsf{T}^\dagger =
\bigl(t_0\mathrm{I}_2 + \vec{t}\cdot \vec{\sigma} \bigl)
\bigl(t_0^*\mathrm{I}_2 + \vec{t}^{\;*}\cdot \vec{\sigma} \bigr)$
which, by the previous identity is
\begin{equation}
\mathcal{S}_{01}(\mathrm{I}_2) = \bigl(t_0 t_0^* + \vec{t}\,\vec{t}^{\;*} \bigr) \mathrm{I}_2 + 
\bigl(t_0^*\vec{t} + t_0 \vec{t}^{\;*} +i \vec{t} \times \vec{t}^{\;*}
\bigr) \cdot \vec{\sigma} \;.
\label{S01_I2}
\end{equation}

Next, it is convenient to evaluate, for a generic Bloch vector $\vec{r}$, the 
operator $\mathsf{T}(\vec{r}\cdot \vec{\sigma})\mathsf{T}^\dagger$, as a useful step toward 
obtaining $\mathcal{S}_{01}(\vec{r}\cdot \vec{\sigma})= 
\mathsf{T}\bigl(U_\xi \vec{r}\cdot \vec{\sigma} \,\bigr)\mathsf{T}^\dagger $. As for 
$\mathcal{S}_{01}(\mathrm{I}_2)$, by performing the (non-commutative) products of Pauli operators, 
we obtain
\begin{eqnarray}
\nonumber
\mathsf{T}(\vec{r}\cdot \vec{\sigma})\mathsf{T}^\dagger &=& 
\bigl[ \bigl(t_0^*\vec{t} + t_0 \vec{t}^{\;*} +i \vec{t}^{\;*}\!\times \vec{t} \;\bigr) \vec{r}\, 
\bigr] \mathrm{I}_2 \\
\label{S01_sr2}
\mbox{} &+& 
\Bigl[ (\vec{t}\,\vec{r}\,)\vec{t}^{\;*} + (\vec{t}^{\;*}\vec{r}\,)\vec{t}
+ \bigl(t_0 t_0^* - \vec{t}\,\vec{t}^{\;*} \bigr)\vec{r}
+ i\bigl(t_0^*\vec{t} - t_0 \vec{t}^{\;*} \bigr) \times \vec{r}\, \Bigr] \cdot \vec{\sigma} \;.
\end{eqnarray}
It can be verified that the Bloch representations in Eqs.~(\ref{S01_I2}) and (\ref{S01_sr2}) 
have real coordinates in the Pauli basis, as it should since $\mathcal{S}_{01}(\rho)$ here is 
Hermitian. For a more concise notation, Eq.~(\ref{S01_sr2}) can be conveniently rewritten as
\begin{equation}
\mathsf{T}(\vec{r}\cdot \vec{\sigma})\mathsf{T}^\dagger =
\bigl( \vec{s}\, \vec{r}\, \bigr) \mathrm{I}_2  + 
A_t \vec{r} \cdot \vec{\sigma} \;.
\label{S01_sr3}
\end{equation}
In Eq.~(\ref{S01_sr3}) has been introduced the $3\times 3$ real matrix $A_t$ function of $t_0$ 
and $\vec{t}$, and defined from Eq.~(\ref{S01_sr2}) by the factor of $\vec{\sigma}$ which is a 
real linear function of $\vec{r}$. In Eq.~(\ref{S01_sr3}), we have also introduced the real 
vector
\begin{equation}
\vec{s}=t_0^*\vec{t} + t_0 \vec{t}^{\;*} +i \vec{t}^{\;*}\!\times \vec{t} = 
2\Re\bigl(t_0^*\vec{t} \;\bigr) +i \vec{t}^{\;*}\!\times \vec{t} 
\label{s_vec}
\end{equation}
of $\mathbbm{R}^3$, which is real since 
$i \vec{t}^{\;*}\!\times \vec{t} = \bigl(i \vec{t}^{\;*}\!\times \vec{t}\;\bigr)^*$ is real. 
We then deduce 
\begin{equation}
\mathcal{S}_{01}(\vec{r}\cdot \vec{\sigma})= 
\mathsf{T}\bigl(U_\xi \vec{r}\cdot \vec{\sigma} \,\bigr)\mathsf{T}^\dagger =
\bigl( \vec{s}\, U_\xi \vec{r}\, \bigr) \mathrm{I}_2  +  A_t U_\xi \vec{r} \cdot \vec{\sigma} \;.
\label{S01_sr4}
\end{equation}

By combining Eqs.~(\ref{S01_I2}) and (\ref{S01_sr4}) we now obtain a characterization for
Eq.~(\ref{S01_T}) in Bloch representation as 
\begin{equation}
\mathcal{S}_{01}(\rho) = \frac{1}{2} 
\bigl( t_0 t_0^* + \vec{t}\,\vec{t}^{\;*} + 
\vec{s}\, U_\xi \vec{r}\, \bigr) \mathrm{I}_2 + \frac{1}{2}
\bigl( t_0^*\vec{t} + t_0 \vec{t}^{\;*} +i \vec{t} \times \vec{t}^{\;*} + A_t U_\xi 
\vec{r}\,\bigr) \cdot \vec{\sigma} \;.
\label{S01_Tbloch}
\end{equation}
Equations~(\ref{S01_Tbloch}) and (\ref{NUNU_ro}) in Bloch representation now provide a 
characterization for the joint two-qubit state $\mathcal{S}(\rho \otimes \rho_c)$ of 
Eq.~(\ref{Sgenqb}) produced by the coherently superposed channel.

We will now examine how quantum measurement on the joint state $\mathcal{S}(\rho \otimes \rho_c)$ 
of Eq.~(\ref{Sgenqb}) produced by the superposed channel, allows us to obtain useful 
information about the unitary $\mathsf{U}_\xi$. This will serve to the analysis of the 
superposed channel for a task of parameter estimation on the unitary $\mathsf{U}_\xi$, the same 
task investigated in \cite{Chapeau21} in a switched channel with indefinite causal order.

\section{Measurement} \label{measur_sec}

The probe qubit prepared in state $\rho$ and the control qubit prepared in state $\rho_c$ get 
coupled by the operation of the coherently superposed quantum channel, and these two qubits 
together terminate in the joint state $\mathcal{S}(\rho \otimes \rho_c)$ of Eq.~(\ref{Sgenqb}). 
To extract information from the superposed channel, a useful strategy, also adopted for instance 
in \cite{Ebler18,Chapeau21}, is to measure the control qubit in the Fourier basis 
$\bigl\{\ket{+}, \ket{-}\bigr\}$ of $\mathcal{H}_2$. The measurement can be described by the two 
measurement operators
$\bigl\{ \mathrm{I}_2 \otimes \ket{+}\bra{+}, \mathrm{I}_2 \otimes \ket{-}\bra{-} \bigr\}$
acting in the Hilbert space $\mathcal{H}_2 \otimes \mathcal{H}_2$ of the probe-control qubit 
pair with state $\mathcal{S}(\rho \otimes \rho_c)$. The measurement randomly projects the control 
qubit either in state $\ket{+}$ or $\ket{-}$, and it leaves the probe qubit in the unnormalized 
conditional state
\begin{equation}
\rho_\pm = {}_c\langle \pm | \mathcal{S}(\rho \otimes \rho_c) | \pm \rangle_c
=\dfrac{1}{2}\mathcal{S}_{00}(\rho) \pm \sqrt{(1-p_c)p_c} \,\mathcal{S}_{01}(\rho) \;,
\label{-S+}
\end{equation}
the partial products involving $\ket{\pm}_c$ being defined on the control qubit. The 
probabilities $P^{\rm con}_\pm$ of the two measurement outcomes are provided by the trace
$P^{\rm con}_\pm =\tr(\rho_\pm )$. From Eq.~(\ref{NUNU_ro}) we have 
$\tr\bigl[\mathcal{S}_{00}(\rho) \bigr] =1$, and by applying the trace on 
Eq.~(\ref{S01_Tbloch}) we obtain
\begin{equation}
P^{\rm con}_\pm =\tr(\rho_\pm ) = \frac{1}{2} \pm \sqrt{(1-p_c)p_c} Q_\xi \;,
\label{tr_ro}
\end{equation}
with the real scalar factor 
\begin{equation}
Q_\xi = \tr\bigl[\mathcal{S}_{01}(\rho) \bigr] = t_0 t_0^* + \vec{t}\,\vec{t}^{\;*} + 
\vec{s}\, U_\xi \vec{r} \;.
\label{Qa1}
\end{equation}
This provides an alternative form for the unnormalized conditional state of Eq.~(\ref{-S+}) as
\begin{equation}
\rho_\pm = \frac{1}{2}
\bigl( P^{\rm con}_\pm \mathrm{I}_2 + \vec{r}_\pm \cdot \vec{\sigma} \,\bigr) \;,
\label{-S+_b}
\end{equation}
with the real vector
\begin{equation}
\vec{r}_\pm =\frac{1}{2}\bigl(A U_\xi +\vec{c}\,\bigr) \vec{r} \pm \sqrt{(1-p_c)p_c} \,
\bigl( t_0^*\vec{t} + t_0 \vec{t}^{\;*} +i \vec{t} \times \vec{t}^{\;*} + A_t U_\xi 
\vec{r}\,\bigr) \;.
\label{r_pm}
\end{equation}

After the measurement of the control qubit, the probe qubit terminates in the (normalized 
conditional) state
\begin{equation}
\rho_\pm^{\rm post}=\dfrac{1}{P^{\rm con}_\pm} \rho_\pm =
\frac{1}{2}\bigl( \mathrm{I}_2 + \vec{r}_\pm^{\rm \,post} \cdot \vec{\sigma} \bigr) \;,
\label{ro_post}
\end{equation}
characterized by the post-measurement Bloch vector 
\begin{equation}
\vec{r}_\pm^{\rm \,post} =\dfrac{\vec{r}_\pm}{P^{\rm con}_\pm} \;,
\label{r_post}
\end{equation}
which is completely known via Eqs.~(\ref{r_pm}) and (\ref{tr_ro}).

An important observation is that, upon measuring the control qubit, the probabilities
$P^{\rm con}_\pm$ of Eq.~(\ref{tr_ro}) governing the measurement outcomes, are in general 
influenced by the unitary $\mathsf{U}_\xi$, via $Q_\xi$ of Eq.~(\ref{Qa1}). The control qubit 
and its measurement can therefore be exploited to extract information about $\mathsf{U}_\xi$. 
It is the probe qubit that interacts with the unitary $\mathsf{U}_\xi$, while the control qubit 
never directly interacts with $\mathsf{U}_\xi$. Nevertheless, the dependence of 
$P^{\rm con}_\pm$ on $\mathsf{U}_\xi$ in Eq.~(\ref{tr_ro}) reveals that it is possible to 
measure the control qubit alone, while discarding the probe qubit, and obtain information about 
the unitary $\mathsf{U}_\xi$. This possibility requires a genuine quantum superposition of two 
distinct channels in Fig.~\ref{figSwi1}, since the dependence of $P^{\rm con}_\pm$ on 
$\mathsf{U}_\xi$, as indicated by Eq.~(\ref{tr_ro}), vanishes for the control at $p_c=0$ or $1$ 
when no superposition exists.

A similar property of a control qubit dependent on $\mathsf{U}_\xi$ was also observed in 
\cite{Chapeau21} for a switched channel with indefinite causal order, although the study in 
\cite{Chapeau21} was restricted to a depolarizing noise $\mathcal{N}(\cdot)$ in 
Fig.~\ref{figUxiN} to enable an analytical characterization of the (more involved)
switched channel. Here with the coherently superposed channel, the property of a
$\mathsf{U}_\xi$-dependent control qubit is generically established in broader conditions, for 
any qubit noise $\mathcal{N}(\cdot)$ in Fig.~\ref{figUxiN} and environment model.
Although the mechanisms are different in the superposed channel here and in the switched 
channel of \cite{Chapeau21}, both realize a coherent control of two channels resulting in a
coupling of the probe and control qubits enabling a $\mathsf{U}_\xi$-dependent control qubit 
that can be measured alone to extract information about $\mathsf{U}_\xi$ sensed by the probe 
qubit alone. This is a remarkable property, not found in conventional metrological techniques
\cite{Giovannetti06,Giovannetti11,Demkowicz14}, where auxiliary inactive qubits that do not 
directly interact with the probed process usually need to be measured coherently with the active 
probing qubits in order to be of some use.

The general description we have obtained of the operation and measurement of the coherently 
superposed qubit channel, will now be applied to a task of phase estimation on the noisy qubit 
unitary $\mathsf{U}_\xi$ engaged in the coherent superposition.

\section{Phase estimation from the superposed channel} \label{phasest_sec}

So we now concentrate, as in \cite{Chapeau21}, on the task of estimating the phase $\xi$ of the 
unitary $\mathsf{U}_\xi$. Phase estimation is an important task of quantum metrology, useful for 
instance in interferometry, magnetometry, frequency standards, atomic clocks, and many other 
high-precision high-sensitivity physical measurements 
\cite{Giovannetti06,Giovannetti11,DAriano98,vanDam07,Chapeau15,Degen17}. The axis $\vec{n}$ of 
the qubit unitary $\mathsf{U}_\xi$ of Eq.~(\ref{Uxi}) is assumed to be known, as fixed by the 
metrological setting, for instance as set by the orientation of the beam splitter of an 
interferometer, along with an unknown phase shift $\xi$ between its two arms that is to be 
estimated. This is a common setting in metrology, where the phase $\xi$ being estimated is 
intended to provide an image of a scalar physical quantity of metrological interest. A 
fundamental tool for assessing and comparing the efficiency of different strategies for 
parameter estimation is provided by the Fisher information, which we now consider.

\subsection{Classical Fisher information for the control qubit} \label{cFisher_sec}

Statistical estimation theory \cite{VanTrees01,Kay93} stipulates that, from measured data 
dependent upon a parameter $\xi$, any conceivable estimator $\widehat{\xi}$ for $\xi$ is endowed 
with a mean-squared error $\langle (\widehat{\xi} -\xi)^2 \rangle$ lower bounded by the 
Cram\'er-Rao inequality, for unbiased as well as for biased estimators with an extended form of 
the inequality \cite{Cover06}. The lower bound to the mean-squared error involves the reciprocal 
of the classical Fisher information $F_c(\xi)$. The larger the Fisher information $F_c(\xi)$, 
the more efficient the estimation can be. The maximum-likelihood estimator \cite{VanTrees01,Kay93} 
is known to achieve the best efficiency dictated by the Cram\'er-Rao lower bound and Fisher 
information $F_c(\xi)$, at least in the asymptotic regime of a large number of independent 
data points. The classical Fisher information $F_c(\xi)$ stands in this respect as a fundamental 
metric quantifying the best achievable efficiency in estimation. 

In the superposed channel, for estimating the phase $\xi$ through the measurement of the control 
qubit displaying the two outcomes characterized by the probabilities $P^{\rm con}_+$ and
$P^{\rm con}_-=1-P^{\rm con}_+$ of Eq.~(\ref{tr_ro}), the classical Fisher information 
\cite{Chapeau16} is
\begin{equation}
F_c^{\rm con}(\xi)= 
\dfrac{(\partial_\xi P^{\rm con}_+)^2}{P^{\rm con}_+} +
\dfrac{(\partial_\xi P^{\rm con}_-)^2}{P^{\rm con}_-} =
\dfrac{(\partial_\xi P^{\rm con}_+)^2}{(1-P^{\rm con}_+)P^{\rm con}_+} \;.
\label{Fc2}
\end{equation}
From Eq.~(\ref{tr_ro}) we have the derivative
\begin{equation}
\partial_\xi P^{\rm con}_+ = \sqrt{(1-p_c)p_c} \partial_\xi Q_\xi =
\sqrt{(1-p_c)p_c} \vec{s}\, \partial_\xi U_\xi \vec{r} \;.
\label{dtr_ro}
\end{equation}
The Fisher information $F_c^{\rm con}(\xi)$ of Eq.~(\ref{Fc2}) follows as
\begin{equation}
F_c^{\rm con}(\xi)= \dfrac{4(1-p_c)p_c (\partial_\xi Q_\xi)^2}{1-4(1-p_c)p_c Q_\xi^2}\;.
\label{Fc2_a}
\end{equation}

The Fisher information $F_c^{\rm con}(\xi)$ of Eq.~(\ref{Fc2_a}) is maximized at $p_c=1/2$, 
which amounts to preparing the control qubit in the state $\ket{\psi_c}=\ket{+}$ realizing an 
equiweighted superposition of the two elementary channels in Fig.~\ref{figSwi1}; we shall stick 
to this favorable condition $p_c=1/2$ in the sequel. The Fisher information of Eq.~(\ref{Fc2_a}) 
then follows as
\begin{equation}
F_c^{\rm con}(\xi)= \dfrac{(\partial_\xi Q_\xi)^2}{1-Q_\xi^2} =\dfrac
{\bigl( \vec{s}\,\partial_\xi U_\xi \vec{r} \,\bigr)^2 }
{ 1 - \bigl( t_0 t_0^* + \vec{t}\,\vec{t}^{\;*} + \vec{s}\, U_\xi \vec{r} \,\bigr)^2  } \;.
\label{Fc2_b}
\end{equation}

For the rotated Bloch vector $\vec{r}_1(\xi)= U_\xi \vec{r}$, a geometric characterization of 
the derivative is provided in \cite{Chapeau16} as
\begin{equation}
\partial_\xi\vec{r}_1(\xi) = \partial_\xi U_\xi \vec{r} = \vec{n} \times \vec{r}_1(\xi) \;.
\label{r1}
\end{equation}
The Fisher information of Eq.~(\ref{Fc2_b}) then becomes
\begin{equation}
F_c^{\rm con}(\xi)= \dfrac{ \bigl[(\vec{s}\times \vec{n}\,)\vec{r}_1 \bigr]^2 }
{ 1 - \bigl( t_0 t_0^* + \vec{t}\,\vec{t}^{\;*} + 
\vec{s} \,\vec{r}_1 \bigr)^2  } \;,
\label{Fc2_c}
\end{equation}
since $\vec{s}\,(\vec{n} \times \vec{r}_1) = (\vec{s}\times \vec{n}\,)\vec{r}_1$.

\begin{figure}[htb]
\centerline{\includegraphics[width=56mm]{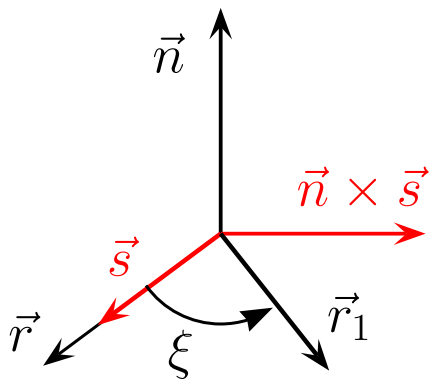}}
\caption[what appears in lof LL p177]
{The vectors in $\mathbbm{R}^3$ from the coherently superposed channel ensuring maximum estimation 
efficiency $F_c^{\rm con}(\xi)$ in Eq.~(\ref{Fc2_c}), with the rotation axis $\vec{n}$ orthogonal 
to $\vec{s}$ of Eq.~(\ref{s_vec}), an input probe of unit Bloch vector 
$\vec{r} \varparallel \vec{s}$, and a rotated Bloch vector $\vec{r}_1(\xi)= U_\xi \vec{r}$ 
remaining orthogonal to $\vec{n}$ for any phase angle $\xi$ to be estimated.
}
\label{figRepz}
\end{figure}

Then Eq.~(\ref{Fc2_c}) shows that a favorable configuration to maximize the Fisher information
$F_c^{\rm con}(\xi)$ is to have an axis $\vec{n}$ orthogonal to the vector $\vec{s}$ of 
$\mathbbm{R}^3$ so as to maximize the magnitude of $\vec{s}\times \vec{n}$, and in addition to 
prepare an input probe with unit Bloch vector $\vec{r}$ also orthogonal to $\vec{n}$ so that the 
rotated Bloch vector $\vec{r}_1(\xi)= U_\xi \vec{r}$ always remains in the plane orthogonal to 
$\vec{n}$ for any angle $\xi$ to estimate. Such an optimal configuration is achieved for instance 
with $\vec{n} \perp \vec{s}$ and $\vec{r} \varparallel \vec{s}$, as depicted in 
Fig.~\ref{figRepz}.

The configuration of Fig.~\ref{figRepz} yields in Eq.~(\ref{Fc2_c}) the Fisher information
\begin{equation}
F_c^{\rm con}(\xi)= \dfrac{\|\vec{s}\,\|^2 \sin^2(\xi)}{1- \bigl[ 
t_0 t_0^* + \vec{t}\,\vec{t}^{\;*} + \|\vec{s}\,\| \, \left|\cos(\xi)\right| \,\bigr]^2 } \;.
\label{Fc2_d}
\end{equation}

We know from Eqs.~(\ref{opTT_1})--(\ref{opTT}) and the example of the Pauli noises, that the real 
vector $\vec{s}$ of Eq.~(\ref{s_vec}) depends on both the noise parameters and the environment 
model via the products $\braket{g | e_j}$, and is independent of the unitary $\mathsf{U}_\xi$
under estimation. Therefore, for any conditions on the noise and 
environment model fixing $\vec{s}$, the arrangement of Fig.~\ref{figRepz} will maximize the 
classical Fisher information at the level of Eq.~(\ref{Fc2_d}), upon measuring the control qubit 
in the Fourier basis for estimating the phase $\xi$. The maximum Fisher information of 
Eq.~(\ref{Fc2_d}) represents a useful reference. It will be achieved in practice if the unitary 
$\mathsf{U}_\xi$ can be operated with an axis $\vec{n} \perp \vec{s}$, for instance by properly 
orienting the interferometer. Otherwise, the efficiency $F_c^{\rm con}(\xi)$ will be reduced 
according to the configuration of $\vec{s}\times \vec{n}$ as dictated by Eq.~(\ref{Fc2_c}).

\subsection{Quantum Fisher information for the control qubit} \label{Qfisher_sec}

To complement the classical Fisher information, further assessment of the estimation performance 
can be obtained by means of the quantum Fisher information \cite{Barndorff00,Paris09}. For a 
$\xi$-dependent qubit state $\rho_\xi$ of Bloch vector $\vec{r}_\xi$, the quantum Fisher 
information relative to the parameter $\xi$ can be expressed \cite{Chapeau16} as
\begin{equation}
F_q(\xi) = \dfrac{\bigl(\vec{r}_\xi\, \partial_\xi\vec{r}_\xi\bigr)^2}{1-\vec{r}_\xi^{\; 2}} +
\bigl( \partial_\xi\vec{r}_\xi \bigr)^2 \;,
\label{Fq_noisyqb}  
\end{equation}
for the general case of a mixed state $\rho_\xi$, while it reduces to
$F_q(\xi) = \bigl( \partial_\xi\vec{r}_\xi \bigr)^2$ for the special case of a pure state 
$\rho_\xi$. The usefulness of $F_q(\xi)$ is that it provides an upper bound to the classical
Fisher information $F_c(\xi)$ attached to any quantum measurement protocol, by imposing 
$F_c(\xi) \le F_q(\xi)$. There might not always exist a fixed $\xi$-independent measurement 
protocol to achieve $F_c(\xi) = F_q(\xi)$, however iterative strategies implementing adaptive 
measurements \cite{Barndorff00,Armen02,Fujiwara06,Brivio10,Tesio11,Okamoto12,Larson17} are 
accessible to achieve $F_c(\xi) = F_q(\xi)$. The quantum Fisher information $F_q(\xi)$ is 
therefore a meaningful metric to characterize the overall best performance for quantum 
estimation.

When the control qubit is measured for estimating the phase $\xi$ while the probe qubit is left 
untouched or unobserved, it is possible to assign a $\xi$-dependent state $\rho^{\rm con}_\xi$ 
to the control qubit by tracing over the probe qubit in the joint probe-control state 
$\mathcal{S}(\rho \otimes \rho_c)$ of Eq.~(\ref{Sgenqb}), yielding
\begin{eqnarray}
\nonumber
\rho^{\rm con}_\xi=\tr_{\rm probe}\bigl[ \mathcal{S}(\rho \otimes \rho_c)\, \bigr]
&=& \tr[\mathcal{S}_{00}(\rho)] \,\bigl[p_c \ket{0_c}\bra{0_c} +(1-p_c)\ket{1_c}\bra{1_c}\,\bigr]\\
\label{Sgenqb_tp1}
&+& \tr[\mathcal{S}_{01}(\rho)] \, \sqrt{(1-p_c)p_c} \bigl(\, \ket{0_c}\bra{1_c} + 
\ket{1_c}\bra{0_c} \,\bigr) \;.
\end{eqnarray}
From Eq.~(\ref{NUNU_ro}) one has $\tr[ \mathcal{S}_{00}(\rho)] =1$, and 
$\tr[ \mathcal{S}_{01}(\rho)] =Q_\xi$ of Eq.~(\ref{Qa1}), so that one finally obtains
\begin{equation}
\rho^{\rm con}_\xi=p_c \ket{0_c}\bra{0_c} + (1-p_c) \ket{1_c}\bra{1_c}  + 
\sqrt{(1-p_c)p_c} Q_\xi \bigl(\, \ket{0_c}\bra{1_c} + \ket{1_c}\bra{0_c}\, \bigr) \;,
\label{Sgenqb_tp2}
\end{equation}
which represents the qubit state characterized by the Bloch vector
$\vec{r}_\xi^{\rm \,con} =\bigl[2\sqrt{(1-p_c)p_c} Q_\xi, 0, 2p_c-1 \bigr]^\top$.
The measurement in the Fourier basis $\bigl\{\ket{+}, \ket{-}\bigr\}$ of the control qubit is 
equivalent to measuring the spin observable $\vec{\omega}_c \cdot \vec{\sigma}$
characterized in $\mathbbm{R}^3$ by the vector $\vec{\omega}_c =\vec{e}_x=[1, 0, 0]^\top$.
When acting on the qubit state $\rho^{\rm con}_\xi$, this spin measurement leads to the
two outcomes $\pm 1$ with the probabilities 
\begin{equation}
P^{\rm con}_\pm =\dfrac{1}{2} \bigl(1 \pm \vec{\omega}_c \vec{r}_\xi^{\rm \,con} \bigr)
\label{Pp+__}
\end{equation}
coinciding with Eq.~(\ref{tr_ro}), while the classical Fisher information is
\begin{equation}
F_c^{\rm \,con}(\xi)= \dfrac{\bigl(\vec{\omega}_c\partial_\xi \vec{r}_\xi^{\rm \,con}\bigr)^2}
{1-\bigl(\vec{\omega}_c \vec{r}_\xi^{\rm \,con}\bigr)^2}
\label{Fc1}
\end{equation}
coinciding with Eq.~(\ref{Fc2}). With the derivative $\partial_\xi \vec{r}_\xi^{\rm \,con} 
=\bigl[2\sqrt{(1-p_c)p_c}\partial_\xi Q_\xi, 0, 0 \bigr]^\top$, Eq.~(\ref{Fq_noisyqb}) provides 
the quantum Fisher information $F_q^{\rm con}(\xi)$ associated with the control qubit as
\begin{eqnarray}
\label{Fq_t1}
F_q^{\rm con}(\xi) 
&=& \dfrac{\bigl[ 4(1-p_c)p_c Q_\xi \partial_\xi Q_\xi \bigr]^2}
{4(1-p_c)p_c \bigl(1-Q^2_\xi\bigr)} 
+ 4(1-p_c)p_c \bigl(\partial_\xi Q_\xi \bigr)^2 \\
&=& 4(1-p_c)p_c \dfrac{\bigl(\partial_\xi Q_\xi \bigr)^2}{1-Q^2_\xi} \;,
\label{Fq_t2}
\end{eqnarray}
forming an upper bound to $F_c^{\rm con}(\xi)$ of Eq.~(\ref{Fc2_a}) as it should.
Just like the classical Fisher information $F_c^{\rm con}(\xi)$ of Eq.~(\ref{Fc2_a}), the 
quantum Fisher information $F_q^{\rm con}(\xi)$ of Eq.~(\ref{Fq_t2}) is maximized at $p_c=1/2$, 
providing an additional motivation to this favorable configuration for preparing the control 
qubit. Then at $p_c=1/2$, one has 
\begin{equation}
F_q^{\rm con}(\xi) = \dfrac{\bigl(\partial_\xi Q_\xi \bigr)^2}{1-Q^2_\xi} \;,
\label{Fq_tp2}
\end{equation}
which coincides with the classical Fisher information $F_c^{\rm con}(\xi)$ of Eqs.~(\ref{Fc2_b}) 
or (\ref{Fc2_c}). In this way, the choice $p_c=1/2$ for the control qubit maximizes the quantum 
Fisher information $F_q^{\rm con}(\xi)$, and in addition achieves 
$F_c^{\rm con}(\xi) =F_q^{\rm con}(\xi)$ for the measurement in the Fourier basis of the control 
qubit, for any $\vec{s}$ and its configuration in relation to the axis $\vec{n}$ and input probe 
$\vec{r}$. In other words, this indicates that there exists a fixed measurement protocol of the 
control qubit, the measurement in the Fourier basis or equivalently via the spin observable 
$\vec{\omega}_c \cdot \vec{\sigma}$ with $\vec{\omega}_c =\vec{e}_x$, able to reach
$F_c^{\rm con}(\xi) =F_q^{\rm con}(\xi)$ --- while as we indicated the existence of such an 
optimal measurement is not granted for all quantum estimation tasks. The choice $p_c=1/2$ and the 
measurement in the Fourier basis for the control qubit, thus constitute the most efficient 
strategy for estimating the phase $\xi$ from the control qubit of the superposed channel; and 
this is equally true for any noise $\mathcal{N}(\cdot)$ and environment model.

\subsection{Coupling by the noise} \label{noiscoup_sec} 

The present analysis reveals the significant property that a non-vanishing noise 
$\mathcal{N}(\cdot)$ in Fig.~\ref{figUxiN} is necessary to couple the control qubit to the 
unitary $\mathsf{U}_\xi$ and enable estimation of the phase $\xi$ from the control qubit of the 
superposed channel. A vanishing noise $\mathcal{N}(\cdot)$ in Fig.~\ref{figUxiN} is represented 
by one trivial Kraus operator $\Lambda_1 = \mathrm{I}_2$ while the other $\Lambda_j$ are the 
null operator. This implies in the Bloch representation of Eq.~(\ref{opTT}) a single nonzero 
coordinate $t_0 =  \braket{g | e_1} $ while $\vec{t}=\vec{0}$. In turn, $\vec{t}=\vec{0}$ 
implies $\vec{s}=\vec{0}$ in Eq.~(\ref{s_vec}) and a factor $Q_\xi$ in Eq.~(\ref{Qa1}) becoming 
independent of the unitary $\mathsf{U}_\xi$. As a result, the measurement probabilities 
$P^{\rm con}_\pm$ of Eq.~(\ref{tr_ro}) are independent of $\mathsf{U}_\xi$. Follows also
$\partial_\xi Q_\xi =0$, so that the Fisher informations $F_c^{\rm con}(\xi)$ in 
Eq.~(\ref{Fc2_a}) and $F_q^{\rm con}(\xi)$ in Eq.~(\ref{Fq_t2}) both vanish, indicating that 
the control qubit is inoperative for estimating the phase $\xi$.

In addition, $\vec{t}=\vec{0}$ leads in Eq.~(\ref{S01_Tbloch}) to
$\mathcal{S}_{01}(\rho) =|t_0|^2 \bigl(\mathrm{I}_2 + U_\xi \vec{r}\, \cdot \vec{\sigma}\bigr)/2
=|t_0|^2 \mathsf{U}_\xi \rho \mathsf{U}_\xi^\dagger$. The joint state of Eq.~(\ref{Sgenqb}) 
factorizes as $\mathcal{S}(\rho \otimes \rho_c) =
\mathsf{U}_\xi \rho \mathsf{U}_\xi^\dagger \otimes \bigl[
p_c \ket{0_c}\bra{0_c} + (1-p_c) \ket{1_c}\bra{1_c} +
|t_0|^2 \sqrt{(1-p_c)p_c} \bigl( \ket{0_c}\bra{1_c} +  \ket{1_c}\bra{0_c} \bigr)\bigr]$
indicating that the probe and control qubits evolve separately, with a control qubit uncoupled 
to $\mathsf{U}_\xi$.  This shows that with no noise 
$\mathcal{N}(\cdot)$ in Fig.~\ref{figUxiN}, although the state $\rho_c$ of the control qubit
is indeed affected, there is however no coupling of the control qubit to the unitary 
$\mathsf{U}_\xi$ which is seen only by the probe qubit. A non-vanishing noise is required to 
couple the control qubit to the unitary $\mathsf{U}_\xi$, via its coupling with the probe qubit 
as described by Eq.~(\ref{Sgenqb}) in the operation of the coherently superposed channel. 
And this is generically true for any type of noise and environment model.

This same property of noise-induced coupling was also observed in the switched channel with 
indefinite causal order studied in \cite{Chapeau21}. This property occurs in the presence of 
two distinguishable channels (1) and (2) superposed in Fig.~\ref{figSwi1} or switched in 
\cite{Chapeau21}. With no noise, the two channels (1) and (2) in Fig.~\ref{figSwi1} 
reduce to two identical and indistinguishable copies of the unitary $\mathsf{U}_\xi$, so that 
the two paths driven by the control qubit are two indistinguishable paths offering limited 
versatility to the probe qubit. By contrast, in the presence of noise, the two channels (1) 
and (2) in Fig.~\ref{figSwi1} involve non-commuting Kraus operators, probabilistically combined, 
and occurring in two independent realizations; this results in two distinct paths driven by 
the control qubit, offering more versatility to the probe qubit, and realizing a quantum
superposition of paths representing a specific nontrivial resource for information processing, as 
also argued in \cite{Ebler18}.

The noise $\mathcal{N}(\cdot)$ in the setting of Fig.~\ref{figUxiN}, if it grows too large, may
be expected also to arrive at some detrimental impact on the estimation performance. We may 
expect in this way an intermediate amount of noise able to maximize the estimation performance 
from the control qubit, that we will explicitly quantify in specific conditions in the sequel. 
Such a beneficial role of noise, here in the superposed channel or in the switched channel of 
\cite{Chapeau21}, more broadly is reminiscent of an effect of stochastic resonance, which 
represents a general phenomenon occurring in various scenarios of  information processing, 
classical \cite{Gammaitoni98,Chapeau99b,McDonnell08,Duan13} or quantum 
\cite{Ting99,Bowen06,Chapeau15c,Gillard17,Gillard19}, and where maximum efficiency is obtained 
at a nonzero optimal amount of noise.

We note also that the coupling of the control qubit to the unitary $\mathsf{U}_\xi$, conveyed by 
the factor $Q_\xi$ in Eq.~(\ref{Qa1}) governing the Fisher informations $F_c^{\rm con}(\xi)$ in 
Eq.~(\ref{Fc2_a}) and $F_q^{\rm con}(\xi)$ of Eq.~(\ref{Fq_t2}), is essentially mediated by the 
vector $\vec{s}$ of Eq.~(\ref{s_vec}). This vector $\vec{s}$ of Eq.~(\ref{s_vec}) can especially 
survive (does not necessarily vanish) in configurations with small $t_0$ and large $\vec{t}$, as 
they would be achieved by noises with Kraus operators applying small probabilistic weight to the 
trivial operator $\mathrm{I}_2$, i.e.\ noises altering the qubit state with large probability. 
Such strong noise configurations, although presumably not optimal, may be expected to preserve 
some capability of the control qubit for phase estimation. In the sequel, in more definite 
conditions, we will further characterize this valuable property of the control qubit of the 
superposed channel, to preserve a capability for phase estimation in the presence of very large 
noise (when conventional estimation techniques fail, as we shall see).

\subsection{Dependence on the noise implementation} \label{implem_sec}

As we have observed, the operation of the coherently superposed channel depends, not only on the
Kraus operators describing each channel (1) and (2) in the superposition, but also on the 
Stinespring representation of each channel, specially via the initial state $\ket{g}$ of the 
environment model in the implementation of each channel. This dependence is encapsulated in the 
two transformation operators $\mathsf{T}_1$ and $\mathsf{T}_2$ of 
Eqs.~(\ref{opT1})--(\ref{opT2}), acting in Eq.~(\ref{S01}). This naturally has an impact when the 
coherently superposed channel analyzed in Section~\ref{switch_sec} is used for a task of phase 
estimation on the noisy qubit unitary of Fig.~\ref{figUxiN}, as we address in the present 
Section~\ref{phasest_sec}. Phase estimation from the control qubit of the superposed channel is 
essentially governed by the two measurement probabilities $P^{\rm con}_\pm$ of Eq.~(\ref{tr_ro}).
These probabilities $P^{\rm con}_\pm$ carry the dependence on the phase $\xi$ via the real scalar 
factor $Q_\xi$ of Eq.~(\ref{Qa1}), which itself depends on the implementation of the noise 
$\mathcal{N}(\cdot )$ via the four Bloch coordinates $(t_0, \vec{t}\,)$ of the operator 
$\mathsf{T}$ in Eq.~(\ref{opTT}). This implies that effective estimation of the phase $\xi$ 
requires some knowledge of the parameters $(t_0, \vec{t}\,)$ related to the noise implementation.
When these parameters cannot be deduced from prior knowledge or assumptions available for the
underlying environment, they represent unwanted unknown parameters; these are rather common in 
estimation theory and are commonly referred to as 
nuisance parameters \cite{Kay93,VanTrees01}. In practice, the presence of these nuisance 
parameters $(t_0, \vec{t}\,)$ in the phase estimation task can be handled in different ways.

First, it can be noted that measurement of the control qubit, via its probabilities
$P^{\rm con}_\pm$ of Eq.~(\ref{tr_ro}), gives direct access to an estimation of the scalar 
factor $Q_\xi$ of Eq.~(\ref{Qa1}). This factor $Q_\xi$ then combines the effect of the unknown 
phase $\xi$ of primary interest, and of the nuisance parameters $(t_0, \vec{t}\,)$. From an 
estimate of $Q_\xi$, the knowledge of $(t_0, \vec{t}\,)$ is necessary if one wants to separate 
and deduce an estimate for the phase $\xi$. However, an estimate of $Q_\xi$ alone may be 
sufficient in practice for some applications. A definite noise $\mathcal{N}(\cdot)$ in the 
setting of Fig.~\ref{figUxiN} is characterized by noise parameters $(t_0, \vec{t}\,)$ that remain 
fixed while 
the phase $\xi$ changes its value. This would be for instance the situation of an interferometer, 
with a definite noise $\mathcal{N}(\cdot)$ characterizing the interferometric setup, while 
operated for estimating the phase shift $\xi$ occurring between its two arms, for metrological 
purposes. The standard modeling of a quantum noise $\mathcal{N}(\cdot)$ by means of a set of 
fixed Kraus operators, as given by Eq.~(\ref{A_Kraus1}), implies the presence of fixed 
environment parameters $(\ket{g}, \mathsf{U_J})$ in the Stinespring implementation of a 
non-unitary channel. This is the same reason that induces fixed noise parameters 
$(t_0, \vec{t}\,)$ for the operation of the coherently superposed channel. In a setup with fixed 
noise parameters $(t_0, \vec{t}\,)$, estimation of $Q_\xi$ alone may be sufficient to track or 
distinguish different values of the phase $\xi$, which would not be known in their absolute 
physical values, but up to some fixed constants determined by $(t_0, \vec{t}\,)$ as a 
characteristic of the metrological instrument. In the estimation of the scalar factor $Q_\xi$, 
the performance is governed by the classical Fisher information $F_c(Q_\xi)$, which relates to 
the Fisher information $F_c(\xi)$ of Eqs.~(\ref{Fc2})--(\ref{Fc2_a}) for the phase $\xi$, by 
$F_c(\xi)=(\partial Q_\xi /\partial \xi)^2 F_c(Q_\xi)$. In particular, $F_c(Q_\xi)$ is maximized 
also at $p_c=1/2$ to reach $F_c(Q_\xi)=1/(1-Q_\xi^2)$. In the same way for the quantum Fisher 
information, $F_q(\xi)=(\partial Q_\xi /\partial \xi)^2 F_q(Q_\xi)$, and $F_q(Q_\xi)$ is also 
maximized at $p_c=1/2$ to reach $F_q(Q_\xi)=1/(1-Q_\xi^2)$, as compared to $F_q(\xi)$ of 
Eq.~(\ref{Fq_tp2}). The performance for estimating $\xi$ or $Q_\xi$ from the control qubit are 
therefore closely related; in particular the scheme operated at $p_c=1/2$ and achieving 
$F_c(Q_\xi)=F_q(Q_\xi)$ is also optimal for estimating $Q_\xi$. An additional interesting feature 
is that the Fisher informations involved (and involving $Q_\xi^2$) never vanish on average, when 
envisaged as averages over the phase $\xi$ or over the noise parameters $(t_0, \vec{t}\,)$, 
indicating that the estimation capabilities of the control qubit are preserved in broad 
configurations of the parameters.

Second, a more thorough approach would be to envisage to estimate the value of the fixed
nuisance parameters $(t_0, \vec{t}\,)$ characterizing the estimation setup. This can be 
accomplished, as for the phase $\xi$, by measuring the control qubit alone in the coherently 
superposed channel. This would again deliver first, via the measurement probabilities 
$P^{\rm con}_\pm$ of Eq.~(\ref{tr_ro}), an estimate for the scalar factor $Q_\xi$ of 
Eq.~(\ref{Qa1}). A calibration procedure could then be envisaged, consisting, from estimates of 
$Q_\xi$ obtained with known test values of the phase $\xi$ arranged for calibration purposes, to 
deduce an estimate for $(t_0, \vec{t}\,)$. These fixed values estimated for the nuisance 
parameters $(t_0, \vec{t}\,)$ would lead to a setup (an interferometer) fit for estimation of 
unknown phases $\xi$. More adequately, estimation is required (and performed from $Q_\xi$ in the 
calibration) rather of the four real scalar parameters $s_0=t_0 t_0^* + \vec{t}\,\vec{t}^{\;*}$ 
and $\vec{s}=[s_x, s_y, s_z]^\top$ from Eq.~(\ref{s_vec}), instead of the four scalar 
$(t_0, \vec{t}\,)$. An access to $\vec{s}$ especially enables one to use the estimation setup in 
the optimal configuration characterized in Fig.~\ref{figRepz}, and $(s_0, \vec{s}\,)$ also 
provide access to the Fisher informations relevant for performance assessment, so that with 
known $(s_0, \vec{s}\,)$ then knowledge of $(t_0, \vec{t}\,)$ is unnecessary. At this occasion 
we can stress this significant property of the coherently superposed channel. The operation of the 
coherently superposed channel is indeed dependent on the environment model implementing each of 
the non-unitary channels in the superposition. However, this dependence is encapsulated in a 
concise transformation operator as in Eqs.~(\ref{opT1})--(\ref{opT2}), 
which, when superposing two identical qubit channels, reduces to a qubit operator as
$\mathsf{T}$ in Eq.~(\ref{opTT}), characterizable with only four scalar parameters
$(t_0, \vec{t}\,)$ or $(s_0, \vec{s}\,)$. This is true independently of the size or 
dimensionality (possibly large) of the environment and of its constitutive details:
only four scalar parameters need be estimated in the calibration procedure.

Alternatively, instead of a prior calibration procedure, direct estimation of the $1+4$ unknown 
parameters $\xi$ and $(s_0, \vec{s}\,)$ can be envisaged also by measuring the control qubit 
alone of the superposed channel. The maximum-likelihood estimator \cite{VanTrees01,Kay93} 
mentioned at the beginning of 
Section~\ref{cFisher_sec}, and matching the performance dictated by the Fisher information, 
typically operates from $L$ independent measurements of the control qubit repeatedly prepared in 
the same conditions. The $L$ measurements, individually governed by the probabilities 
$P^{\rm con}_\pm$ of Eq.~(\ref{tr_ro}), deliver $L_+$ outcomes projecting on $\ket{+}$ and 
$L-L_+$ on $\ket{-}$, with the integer $L_+ \in \{0, 1, \cdots, L \}$. From the binomial 
probability distribution of the $L$ outcomes follows the log-likelihood
$\mathcal{L}=L_+\log\bigl(P^{\rm con}_+ \bigr)+(1-L_+)\log\bigl(1-P^{\rm con}_+ \bigr)$.
From the measured data $(L_+, L-L_+)$, the maximum-likelihood estimator for the multiparameter
set $(\xi, s_0, \vec{s}\,)$ follows as the solution to
$\text{arg\,max}_{(\xi, s_0, \vec{s}\,)} \mathcal{L}$; while in the presence of known 
$(s_0, \vec{s}\,)$ through prior calibration, the maximum-likelihood estimator for the single 
phase parameter $\xi$ is the solution to $\text{arg\,max}_{\xi} \mathcal{L}$. Such a direct 
estimation of a whole set $(\xi, s_0, \vec{s}\,)$ of unknown parameters can also be realized 
recursively, in an adaptive manner, where especially the estimation of $\vec{s}$ is progressively 
updated to bring the setup in the optimal configuration of Fig.~\ref{figRepz} maximizing the 
performance. Such adaptive strategies have proven useful in other areas of quantum estimation and 
allow one to reach or to come close to optimality 
\cite{Barndorff00,Armen02,Fujiwara06,Brivio10,Tesio11,Okamoto12,Larson17}.

\subsection{Analysis with Pauli noise} \label{Paulin_sec} 

For further, more definite, illustration, we consider the interesting case already mentioned in 
Section~\ref{qbswitch_sec} where $\mathcal{N}(\cdot)$ is a generic Pauli noise 
\cite{Nielsen00,Wilde17} with the four Kraus operators 
$\{\Lambda_j \}= \bigl\{\sqrt{p_0}\mathrm{I}_2, \sqrt{p_x}\sigma_x, \allowbreak
\sqrt{p_y}\sigma_y, \sqrt{p_z}\sigma_z \bigr\}$ and $\{p_j \}$ a probability distribution. 
For definiteness, we consider that the Pauli noise $\mathcal{N}(\cdot)$ 
is associated with an environment model starting in an unbiased initial state $\ket{g}$ 
satisfying $\braket{g | e_j} =1/2$ for all $j=1$ to $4$. This represents an environment that 
does not specially favor any one of the underlying Kraus operators, but on the contrary that 
treats them in an even, equally weighted, manner. This is a reasonable configuration for an 
environment initially uncorrelated with the signal-control compound, realizing some kind of 
least biased or maximum entropy environment. This is also the choice that is considered for 
instance in the superposed channels investigated for information communication in 
\cite{Abbott20,Loizeau20}. Such assumption on the noise constitution dispenses us from handling 
nuisance parameters as examined in Section~\ref{implem_sec}. For the Bloch representation of 
$\mathsf{T}$ in Eq.~(\ref{opTT}) we then have the coordinates $t_0=\sqrt{p_0}/2$ and 
$\vec{t}=\bigl[\sqrt{p_x}, \sqrt{p_y}, \sqrt{p_z}\,\bigr]^\top /2$, so that 
$t_0 t_0^* + \vec{t}\,\vec{t}^{\;*}=1/4$, and 
$\vec{s}=2t_0\vec{t}=\sqrt{p_0}\bigl[\sqrt{p_x}, \sqrt{p_y}, \sqrt{p_z}\,\bigr]^\top /2$ in 
Eq.~(\ref{s_vec}) yielding  $\|\vec{s}\,\|^2=(1-p_0)p_0/4$. The Fisher information from 
Eqs.~(\ref{Fc2_d}) and (\ref{Fq_tp2}) follows as 
\begin{equation}
F_c^{\rm con}(\xi)= F_q^{\rm con}(\xi)=\dfrac{(1-p_0)p_0 \sin^2(\xi)}
{ 4 - \biggl[ \frac{1}{2} + \sqrt{(1-p_0)p_0} \left|\cos(\xi)\right| \biggr]^2 } \;.
\label{Fc2_Pauli}
\end{equation}

We observe that Eq.~(\ref{Fc2_Pauli}) now enables an explicit quantification of the intermediate 
amount of noise anticipated in Section~\ref{noiscoup_sec} to maximize the performance of the 
control qubit. The estimation performance assessed by the Fisher information 
$F_c^{\rm con}(\xi)=F_q^{\rm con}(\xi)$ of Eq.~(\ref{Fc2_Pauli}) takes its maximum for a 
probability $p_0= 1/2$, which is the situation where the noise maintains unaltered the qubit 
state (by applying the trivial Kraus operator $\Lambda_1=\sqrt{p_0}\mathrm{I}_2$) with 
probability $1/2$. This behavior holds in the same way for any type of Pauli noise, irrespective 
of the other probabilities $\{p_x, p_y, p_z\}$ provided $p_x+p_y+p_z =1-p_0=1/2$. This remarkable 
property is obtained, from Eq.~(\ref{Fc2_c}), with a rotation axis $\vec{n}$ orthogonal to 
$\vec{s}$, with $\vec{s}$ itself depending on $\{p_x, p_y, p_z\}$; otherwise, with $\vec{n}$ and 
$\vec{s}$ in another configuration, the value of $F_c^{\rm con}(\xi)= F_q^{\rm con}(\xi)$ given 
by Eq.~(\ref{Fc2_c}) can be expected to be reduced relative to Eq.~(\ref{Fc2_Pauli}) and 
dependent on the noise probabilities $\{p_x, p_y, p_z\}$. 

It can also be noted that the Fisher information $F_c^{\rm con}(\xi)=F_q^{\rm con}(\xi)$ of 
Eq.~(\ref{Fc2_Pauli}) bears an explicit dependence on the phase $\xi$, and this is also the rule 
in the more general conditions of Eqs.~(\ref{Fc2_d}) or (\ref{Fc2_c}). This is a common 
property, often observed for quantum phase estimation in the presence of noise, and implying a 
performance varying according to the range of the phase $\xi$ being estimated. A measurement 
result depending on $\xi$ is necessary to enable estimation of $\xi$ from such measurement. 
Commonly this entails also a measurement performance depending on $\xi$. For a performance 
assessment circumventing this dependence in $\xi$, it can be meaningful to consider the averaged 
Fisher information $\overline{F}_c = \int_0^{2\pi}F_c(\xi) d\xi/(2\pi)$ reflecting the average 
performance for values of the phase angle $\xi$ uniformly covering the interval $[0, 2\pi )$. 
The averaged Fisher information $\overline{F}_c$ also determines (via its reciprocal) a 
fundamental lower bound to the mean-squared estimation error in Bayesian estimation 
\cite{VanTrees01}. Here, for the Fisher information $F_c^{\rm con}(\xi)=F_q^{\rm con}(\xi)$ in 
Eq.~(\ref{Fc2_Pauli}) of the control qubit, the integral over $\xi$ can be worked out explicitly 
to give
\begin{equation}
\overline{F}_c^{\rm con} (\xi)=\overline{F}_q^{\rm con} (\xi)= 1- \frac{1}{2\pi} \left[
\sqrt{9-\beta^2}\arctan\Biggl( \sqrt{\frac{3+\beta}{3-\beta}} \,\Biggr) +
\sqrt{25-\beta^2}\arctan\Biggl( \sqrt{\frac{5-\beta}{5+\beta}} \,\Biggr) \right] 
\label{Fc_av}
\end{equation}  
(where $\xi$ is kept in the notation after the average to indicate the Fisher information is 
relative to the parameter $\xi$), with the noise factor 
\begin{equation}
\beta=2\sqrt{(1-p_0)p_0} \;.
\label{beta1}
\end{equation}  
And Eq.~(\ref{Fc_av}) will be useful for quantitative performance comparison to come.

\section{Performance comparison} \label{standqb_sec}

A meaningful reference for comparison is the Fisher information for estimating the phase $\xi$ 
from a probing qubit that would directly interact with the noisy unitary channel of 
Fig.~\ref{figUxiN} in a single pass, with no coherent superposition of two copies of the channel. 
For the probing qubit prepared in the state $\rho$ of Eq.~(\ref{roBloch}), one pass through the 
channel of Fig.~\ref{figUxiN} is described by the quantum operation of Eq.~(\ref{NUxi_ro}), and 
it leaves the qubit in a state characterized by the Bloch vector
\begin{equation}
\vec{r}_\xi= A U_\xi \vec{r} + \vec{c} =A \vec{r}_1(\xi) +\vec{c}\;.
\label{r1xi}
\end{equation}
With the derivative $\partial_\xi \vec{r}_\xi= A \partial_\xi \vec{r}_1(\xi)$ characterized 
by Eq.~(\ref{r1}), the quantum Fisher information is obtained from Eq.~(\ref{Fq_noisyqb}) as
\begin{equation}
F_q(\xi) = \dfrac{\bigl[(A\vec{r}_1+\vec{c}\,)A (\vec{n}\times \vec{r}_1) \bigr]^2}
{1- (A\vec{r}_1 +\vec{c}\,)^2 } + \bigl[A (\vec{n}\times \vec{r}_1) \bigr]^2 \;.
\label{Fq_stand1}  
\end{equation}
When this standard probing qubit is measured, as the control qubit of the superposed channel
by means of a spin observable $\vec{\omega} \cdot \vec{\sigma}$ characterized by a unit Bloch 
vector $\vec{\omega} \in \mathbbm{R}^3$, the ensuing classical Fisher information is obtained 
from Eq.~(\ref{Fc1}) as
\begin{equation}
F_c(\xi)= \dfrac{\bigl(\vec{\omega}\,\partial_\xi \vec{r}_\xi\bigr)^2}
{1-\bigl(\vec{\omega}\, \vec{r}_\xi\bigr)^2} =
\dfrac{\bigl[\vec{\omega}\, A (\vec{n}\times \vec{r}_1) \bigr]^2}
{1- \bigl[\vec{\omega}\, (A\vec{r}_1 +\vec{c}\,)\bigr]^2 }  \;.
\label{Fc_stand1}
\end{equation}

It is then meaningful to compare the performance of such a standard probing qubit with the
performance $F_c^{\rm con}(\xi)=F_q^{\rm con}(\xi)$ of Eq.~(\ref{Fc2_c}) achieved at $p_c=1/2$ 
by the control qubit of the coherently superposed channel measured in the Fourier basis with 
$\vec{\omega}_c =\vec{e}_x$.

For specifying the comparison in the conditions of Section~\ref{Paulin_sec}, for the class of 
Pauli noises having $\vec{c}\equiv\vec{0}$, it is shown in \cite{Chapeau15} that a 
favorable configuration to maximize the quantum Fisher information $F_q(\xi)$ of 
Eq.~(\ref{Fq_stand1}) is when the rotated Bloch vector $\vec{r}_1(\xi)$ is a unit vector
orthogonal to the rotation axis $\vec{n}$, which is obtained by a pure input probe with a unit 
Bloch vector $\vec{r}$ also orthogonal to $\vec{n}$. This choice for the input probe maintains 
the rotated Bloch vector $\vec{r}_1(\xi)$ orthogonal to $\vec{n}$ for any phase angle $\xi$.
With $\vec{r}$ and $\vec{r}_1(\xi)$ orthogonal to $\vec{n}$, there is however no fixed 
measurement vector $\vec{\omega}$ generally enabling the classical Fisher information $F_c(\xi)$ 
of Eq.~(\ref{Fc_stand1}) to reach the quantum Fisher information $F_q(\xi)$ of 
Eq.~(\ref{Fq_stand1}), even for a fixed type of Pauli noise. In general, achieving 
$F_c(\xi)=F_q(\xi)$ with the standard probing qubit, would require a vector $\vec{\omega}$ 
dependent on the noise matrix $A$ but also on the unknown phase $\xi$ under estimation, as it 
results from \cite{Chapeau15,Chapeau16}. By contrast, with the superposed channel, 
$F_c^{\rm con}(\xi)=F_q^{\rm con}(\xi)$ is achievable by $\vec{\omega}_c =\vec{e}_x$ for 
measuring the control qubit, uniformly for any noise and environment model, as we observed in 
Section~\ref{Qfisher_sec}.

For further quantitative comparison, it is interesting to work out the case of the qubit 
depolarizing noise, an important instance of Pauli noise \cite{Nielsen00,Wilde17}. The 
depolarizing noise applies any one of the three non-trivial Pauli operators with equal 
probabilities $p_x=p_y=p_z=p/3$, while it leaves the qubit state $\rho$ unchanged (or applies
$\mathrm{I}_2$) with the probability $p_0=1-p$. In Eq.~(\ref{Pauli1}), along with 
$\vec{c}\equiv\vec{0}$, the resulting matrix $A$ is the isotropic contraction matrix 
$A=\alpha I_3$ with $\alpha =1-4p/3$ a scalar contraction factor, implementing an isotropic 
contraction of the qubit Bloch vector $\vec{r} \mapsto \alpha \vec{r}$. Equivalently, the action 
of the depolarizing noise can be described as
\begin{equation}
\mathcal{N}(\rho)= \alpha\rho + (1-\alpha) \frac{\mathrm{I}_2}{2} \;,
\label{depol2}
\end{equation}
indicating that with the probability $1-\alpha$, the noise replaces the qubit state $\rho$ by
the maximally mixed state $\mathrm{I}_2/2$; at the maximum contraction when $\alpha =0$, the 
quantum state is forced to $\mathrm{I}_2/2$ with probability $1$ and the qubit becomes completely 
depolarized. The depolarizing noise is an important noise model often considered in quantum 
information \cite{Nielsen00,Wilde17}. It is a Pauli noise with no invariant subspace, unlike the 
bit-flip or phase-flip noises, and in this respect it 
represents in some sense a worse-case noise and as such a conservative reference. 

With depolarizing noise, for the standard probing qubit, the quantum Fisher information of 
Eq.~(\ref{Fq_stand1}) becomes
\begin{equation}
F_q(\xi) = \alpha^2 \bigl(\vec{n}\times \vec{r}_1 \bigr)^2 
= \alpha^2 \bigl(\vec{n}\times \vec{r}\, \bigr)^2 \;,
\label{Fq_stand2}  
\end{equation}
and the classical Fisher information of Eq.~(\ref{Fc_stand1}), 
\begin{equation}
F_c(\xi)= \dfrac{\alpha^2 \bigl[\vec{\omega}\, (\vec{n}\times \vec{r}_1) \bigr]^2}
{1- \alpha^2(\vec{\omega}\,\vec{r}_1)^2 } \;.
\label{Fc_stand2}
\end{equation}
In the favorable configuration where $\vec{r}$ and $\vec{r}_1(\xi)$ are in the plane orthogonal 
to the rotation axis $\vec{n}$, the quantum Fisher information of Eq.~(\ref{Fq_stand2}) is 
maximized at $F_q^{\rm max}(\xi)=\alpha^2$ for any phase angle $\xi$. However, as indicated, 
there is no fixed measurement vector $\vec{\omega}$ enabling $F_c(\xi)$ of Eq.~(\ref{Fc_stand2})
to reach $F_q^{\rm max}(\xi)=\alpha^2$, for any phase $\xi$. This would require an $\vec{\omega}$
orthogonal to the rotated Bloch vector $\vec{r}_1(\xi)$ for any $\xi$, which is not realizable 
since $\xi$ is unknown. Alternatively, one can select $\vec{\omega}$ to coincide with the pure 
input probe $\vec{r}$, so that at $\vec{\omega}=\vec{r} \perp \vec{n}$, the classical Fisher 
information of Eq.~(\ref{Fc_stand2}) becomes
\begin{equation}
F_c(\xi)= \dfrac{\alpha^2 \sin^2(\xi)}{1- \alpha^2\cos^2(\xi) } \le F_q(\xi) = \alpha^2 \;.
\label{Fc_stand3}
\end{equation}
This $F_c(\xi)$ is to be compared with $F_c^{\rm con}(\xi)= F_q^{\rm con}(\xi)$ of
Eq.~(\ref{Fc2_Pauli}) with for the depolarizing noise the factor 
$(1-p_0)p_0 = 3(1+3\alpha)(1-\alpha)/16$, or the equivalent form of Eq.~(\ref{Fc2_Pauli}) as
\begin{equation}
F_c^{\rm con}(\xi)=F_q^{\rm con}(\xi)= \frac{\beta^2 \sin^2(\xi)}
{ 16 - \bigl[ 1+\beta \left|\cos(\xi)\right| \,\bigr]^2 } \;,
\label{Fc2_e}
\end{equation}
with the factor $\beta$ of Eq.~(\ref{beta1}) expressed as a function of the contraction factor 
$\alpha$ as
\begin{equation}
\beta=\frac{\sqrt{3}}{2}\sqrt{(1-\alpha )(1+3\alpha )} \;.
\label{beta2}
\end{equation} 
The performance of Eq.~(\ref{Fc2_e}) is achieved in the superposed channel by the control qubit 
in the optimal configuration of Fig.~\ref{figRepz} with a vector $\vec{s}$ in Eq.~(\ref{s_vec}) 
which for the depolarizing noise is 
$\vec{s}= \beta/\bigl(4\sqrt{3}\bigr) [1, 1, 1]^\top = \beta/\bigl(4\sqrt{3}\bigr)\vec{1}$.
This is obtained with an axis $\vec{n} \perp \vec{1}$ and an input probe 
$\vec{r} \varparallel \vec{1}$, uniformly for any level $\alpha$ of the depolarizing noise.
By contrast, as we indicated, the standard probing qubit in general cannot reach 
$F_c(\xi)= F_q(\xi)$ in Eq.~(\ref{Fc_stand3}) or (\ref{Fc_stand2}).

Then the Fisher informations of Eqs.~(\ref{Fc_stand3}) and (\ref{Fc2_e}) offer a meaningful
basis for quantitative comparison. It can be noted that all of them depend on the value of the 
phase $\xi$ under estimation. As indicated at the occasion of Eq.~(\ref{Fc2_Pauli}), to 
circumvent the $\xi$-dependent performance it is convenient to consider the phase-averaged 
Fisher information for a phase $\xi$ uniformly covering the interval $[0, 2\pi )$. For the 
control qubit of the coherently superposed channel, such phase averaging of Eq.~(\ref{Fc2_e}) 
leads to $\overline{F}_c^{\rm con}(\xi)=\overline{F}_q^{\rm con}(\xi)$ of Eq.~(\ref{Fc_av}) 
with $\beta$ from Eq.~(\ref{beta2}). Meanwhile, for the standard probing qubit the phase 
averaging of Eq.~(\ref{Fc_stand3}) yields
\begin{equation}
\overline{F}_c(\xi) = 1- \sqrt{1-\alpha^2} \le \overline{F}_q(\xi) =\alpha^2 \;.
\label{Fc_av_stand}
\end{equation}  

For comparison, we also consider the phase-averaged Fisher information achieved by the control 
qubit of a switched channel with indefinite causal order, as analyzed in \cite{Chapeau21},
which is given by
\begin{equation}
\overline{F}_c^{\rm swi}(\xi)=\overline{F}_q^{\rm swi}(\xi) = 
1-\frac{\sqrt{3}}{8}(1-\alpha) \sqrt{(1-\alpha)(3+5\alpha)} 
 - \frac{1}{8} \sqrt{(5+6\alpha-3\alpha^2)(5-2\alpha+5\alpha^2)} \;,
\label{Fc_av_swi}
\end{equation}  
reproducing Eq.~(44) of \cite{Chapeau21}.

\begin{figure}[htb]
\centerline{\includegraphics[width=84mm]{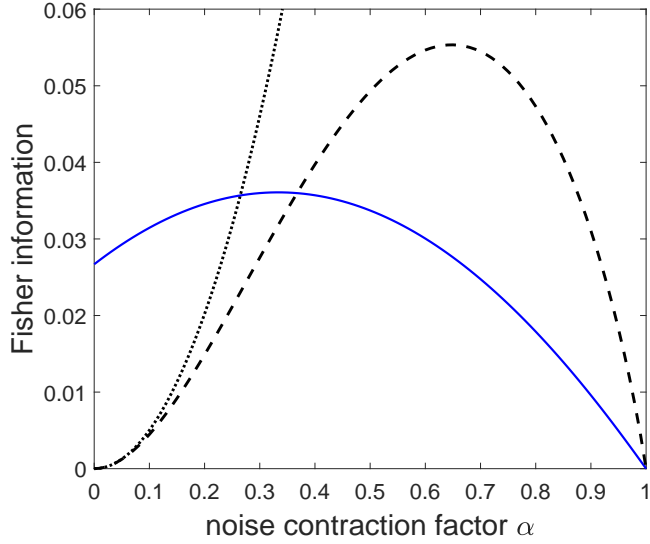}}
\caption[what appears in lof LL p177]
{Phase-averaged Fisher information $\overline{F}_c^{\rm con}(\xi)=\overline{F}_q^{\rm con}(\xi)$ 
of Eq.~(\ref{Fc_av}) via Eq.~(\ref{beta2}) for the control qubit of the coherently superposed 
channel (solid line), $\overline{F}_c(\xi)$ of Eq.~(\ref{Fc_av_stand}) for the standard 
probing qubit (dotted line), 
and $\overline{F}_c^{\rm swi}(\xi)=\overline{F}_q^{\rm swi}(\xi)$ of Eq.~(\ref{Fc_av_swi}) for 
the control qubit of the switched channel with indefinite order of \cite{Chapeau21} (dashed line),
as a function of the contraction factor $\alpha$ of the depolarizing noise of Eq.~(\ref{depol2}).
}
\label{figFcav1}
\end{figure}

The three phase-averaged Fisher informations are compared in Fig.~\ref{figFcav1}, as a function
of the level of the depolarizing noise quantified by its contraction factor $\alpha$.

\smallbreak

As visible in Fig.~\ref{figFcav1}, for most noise levels $\alpha$, the straightforward approach 
of the standard probing qubit is more efficient, with a larger Fisher information 
$\overline{F}_c(\xi)$. At vanishing noise, when $\alpha \rightarrow 1$, the Fisher information 
$\overline{F}_c(\xi)$ of Eq.~(\ref{Fc_av_stand}) goes to $1$ and reaches its upper bound 
$\overline{F}_q(\xi)=\alpha^2 \rightarrow 1$ where the estimation efficiency of the standard 
probing qubit is maximal. By contrast, the control qubit of the superposed channel, at vanishing 
noise when $\alpha \rightarrow 1$, as anticipated in Section~\ref{noiscoup_sec}, becomes 
inoperative for estimating the phase $\xi$, as marked by a vanishing Fisher information 
$\overline{F}_c^{\rm con}(\xi)=\overline{F}_q^{\rm con}(\xi)$ in Fig.~\ref{figFcav1}.
In addition, Fig.~\ref{figFcav1} also shows the existence of an optimal level of noise where 
the efficiency of the control qubit of the superposed channel is maximized, as also anticipated 
in Section~\ref{noiscoup_sec}. This is marked in Fig.~\ref{figFcav1} by a Fisher information 
$\overline{F}_c^{\rm con}(\xi)=\overline{F}_q^{\rm con}(\xi)$ culminating at a maximum for 
$\alpha \approx 0.34$.

An interesting feature shown in Fig.~\ref{figFcav1} is that when the noise becomes fully 
depolarizing, at $\alpha \rightarrow 0$, the Fisher information 
$\overline{F}_c^{\rm con}(\xi)=\overline{F}_q^{\rm con}(\xi)$ of the control qubit of the 
superposed channel does not go to zero, 
while $\overline{F}_c(\xi)$ and $\overline{F}_q(\xi)$ vanish for the standard probing 
qubit. At $\alpha \rightarrow 0$, the standard probing qubit ruled by Eq.~(\ref{NUxi_ro}) with 
$A=\alpha I_3$ and $\vec{c}\equiv \vec{0}$, experiences the input--output transformation 
$\rho \mapsto \mathrm{I}_2 /2$, producing a completely depolarized output state 
$\mathrm{I}_2 /2$ insensitive to the phase $\xi$, and therefore inoperative for its estimation, 
as expressed by the vanishing Fisher informations $\overline{F}_c(\xi)$ and 
$\overline{F}_q(\xi)$ in Eq.~(\ref{Fc_av_stand}) and visible in Fig.~\ref{figFcav1}.
By contrast, via the coupling interaction taking place in the coherently superposed channel,
the output state $\rho^{\rm con}_\xi$ of Eq.~(\ref{Sgenqb_tp2}) for the control qubit remains 
sensitive to the phase $\xi$, even at $\alpha \rightarrow 0$, as expressed by the non-vanishing 
Fisher information $\overline{F}_c^{\rm con}(\xi)=\overline{F}_q^{\rm con}(\xi)$ in 
Fig.~\ref{figFcav1}. This is a significant difference, with a fully depolarizing noise in the 
noisy unitary channel of Fig.~\ref{figUxiN}, the standard probing qubit is unable to deliver 
information about the phase $\xi$; by contrast, the control qubit of the superposed channel of
Fig.~\ref{figSwi1}, through the more elaborate coupling interaction analyzed here, remains 
operative for estimating $\xi$. This is a manifestation of the beneficial role of the noise in 
coupling the control qubit to the unitary $\mathsf{U}_\xi$ in the superposed channel, as 
discussed in Section~\ref{noiscoup_sec}. Accordingly, at strong noise, when the contraction 
factor $\alpha \lesssim 0.26$ in Fig.~\ref{figFcav1}, the Fisher information, and therefore the 
efficiency for phase estimation, of the control qubit of the superposed channel, remain 
superior to those of the standard probing qubit.  

In addition, the comparison with the switched channel with indefinite order of \cite{Chapeau21} 
shows in Fig.~\ref{figFcav1} a switched channel with a Fisher information
$\overline{F}_c^{\rm swi}(\xi)=\overline{F}_q^{\rm swi}(\xi)$ of Eq.~(\ref{Fc_av_swi}) also
undergoing a non-monotonic evolution, as 
$\overline{F}_c^{\rm con}(\xi)=\overline{F}_q^{\rm con}(\xi)$ of Eq.~(\ref{Fc_av}) for the
superposed channel, yet culminating at a maximum for a different noise level $\alpha \approx 0.65$.
Also, for the noise contraction factor above the crossover $\alpha \approx 0.37$, the performance 
of the switched channel remains larger than that of the superposed channel. The operating 
mechanisms of the two channels are significantly different, but the larger performance can be 
related to the fact that the elementary path in the switched channel of \cite{Chapeau21} includes 
two passes across the sensed unitary $\mathsf{U}_\xi$, when the superposed channel here, in 
comparison, includes only one. A significant observation is that, at large noise when 
$\alpha \rightarrow 0$, the switched channel also exhibits a vanishing performance 
$\overline{F}_c^{\rm swi}(\xi)=\overline{F}_q^{\rm swi}(\xi) \rightarrow 0$, and becomes 
inoperative for phase estimation, just like the standard probing qubit; while, as we already
mentioned, the superposed channel remains operative for estimation in this large-noise limit.
This is a remarkable benefit observed with the coherently superposed channel, to maintain 
the possibility of phase estimation, even when the unitary $\mathsf{U}_\xi$ in Fig.~\ref{figUxiN}
is buried in a fully depolarizing noise. No other estimation techniques are known to share this
ability.

Another important difference is that the coherently superposed channel here, in order for its 
control qubit to reach the optimized performance of Eq.~(\ref{Fc2_d}) or (\ref{Fc_av}), requires 
a pure input probe with a unit Bloch vector $\vec{r}$ prepared in the configuration of 
Fig.~\ref{figRepz}. By contrast, the switched channel of \cite{Chapeau21} exhibits a control
qubit reaching the performance conveyed by Eq.~(\ref{Fc_av_swi}) independently of the
preparation $\vec{r}$ of the input probe qubit. Especially, even when $\vec{r}=\vec{0}$ with a 
completely depolarized input probe qubit, the control qubit of the switched channel in 
\cite{Chapeau21} maintains its capability for phase estimation, with the performance of 
Eq.~(\ref{Fc_av_swi}), but this capability disappears in the presence of a fully depolarizing 
noise $\mathcal{N}(\cdot)$ in Fig.~\ref{figUxiN}. So the superposed channel here requires a pure 
input probe qubit optimized as in Fig.~\ref{figRepz}, but its capability for phase estimation 
persists even with a fully depolarizing noise $\mathcal{N}(\cdot)$; by contrast, the switched 
channel of \cite{Chapeau21} remains operative for phase estimation even with a completely 
depolarized input probe qubit, but its capability vanishes with a fully depolarizing noise 
$\mathcal{N}(\cdot)$. These are two distinct properties, stemming from the different mechanisms 
ruling the superposed or the switched channels, but that are not accessible with conventional
estimation techniques. These unconventional properties can be related to the results in quantum 
communication of \cite{Ebler18,Loizeau20,Abbott20}, also concerning completely depolarizing 
channels, which recover ability for information communication when they are combined into a 
switched channel with indefinite order in \cite{Ebler18,Loizeau20} or into a coherently 
superposed channel in \cite{Abbott20,Loizeau20}. Beyond quantum communication in 
\cite{Ebler18,Loizeau20,Abbott20}, the possibility of recovering from fully depolarizing 
conditions, through coherent control on quantum channels, under different forms here and in 
\cite{Chapeau21}, is extended to the fundamental metrological task of quantum phase estimation.

\bigbreak 
To further appreciate the beneficial role of noise in the operation of the coherently
superposed channel, we can evaluate, as in \cite{Loizeau20}, an index of non-commutativity
$\rchi_\text{NC}$ of the underlying Kraus operators, by means of a cumulated trace distance 
between their pairwise commutators, and defined in the conditions of Section~\ref{qbswitch_sec}
as $\rchi_\text{NC}=\sum_{j,k} 
\tr\bigl( [\mathsf{K}_j, \mathsf{K}_k]^\dagger [\mathsf{K}_j, \mathsf{K}_k] \bigr)$.
In general this non-commutativity index $\rchi_\text{NC}$ can be expected to depend on the
parameters of the underlying Kraus operators, in particular here with the channels of 
Fig.~\ref{figUxiN} on the parameters of the unitary $\mathsf{U}_\xi$ and the noise 
$\mathcal{N}(\cdot)$. For illustration, for the channels with depolarizing noise studied in 
Section~\ref{standqb_sec} and Fig.~\ref{figFcav1}, the non-commutativity index evaluates to 
\begin{equation}
\rchi_\text{NC} =8(1-\alpha)\alpha \sin^2(\xi/2) + 3(1-\alpha)^2 \;,
\label{NCindex}
\end{equation}  
which indeed depends on the unitary $\mathsf{U}_\xi$ via its phase $\xi$ and on the noise
via its contraction factor $\alpha$. We observe, consistently, that the non-commutativity index 
$\rchi_\text{NC}$ of Eq.~(\ref{NCindex}) vanishes at $\alpha =1$ when there is no noise and the 
Kraus operators commute. On the contrary, $\rchi_\text{NC}$ in Eq.~(\ref{NCindex}) tends to be 
large at strong noise when $\alpha \rightarrow 0$, showing that the noise is an essential 
ingredient to make the Kraus operators non-commuting and favor the probe-control coupling in the 
superposed channel, as discussed in Section~\ref{noiscoup_sec}. The phase-averaged index 
$\overline{\rchi}_\text{NC} =4(1-\alpha)\alpha + 3(1-\alpha)^2=(1-\alpha)(3+\alpha)$ culminates 
at $\overline{\rchi}_\text{NC}=3$ at maximum noise when $\alpha =0$ and monotonically decays to 
zero as $\alpha \rightarrow 1$. In these conditions, maximum noise achieves maximum 
non-commutativity of the Kraus operators. However, as discussed in Section~\ref{noiscoup_sec}, 
stronger noise, which favors non-commutativity and coupling, at some point arrives at a 
detrimental impact on the phase estimation efficiency, whence the intermediate level of noise 
as a compromise to maximize the performance of phase estimation from the control qubit in the 
superposed channel, as reflected for instance in Fig.~\ref{figFcav1}.

\bigbreak
A non-vanishing Fisher information $F_c^{\rm con}(\xi)=F_q^{\rm con}(\xi)$ observed for
the control qubit of the superposed channel in the presence of a fully depolarizing noise
at $\alpha =0$, is generically preserved with any underlying environment model.
Instead of the unbiased environment characterized by $\braket{g | e_j} =1/2$ for $j=1$ to 
$4$ chosen in Section~\ref{Paulin_sec}, an arbitrary environment underlying the depolarizing
noise of Eq.~(\ref{depol2}) can be handled with in Eq.~(\ref{opTT}) the general coordinates
$t_0 = \braket{g | e_1}\!\sqrt{p_0} =\braket{g | e_1}\!\sqrt{1+3\alpha}/2$ and
$\vec{t}=\vec{e}\sqrt{(1-p_0)/3}=\vec{e}\sqrt{1-\alpha}/2$ with the vector
$\vec{e}=\bigl[\braket{g | e_2}, \braket{g | e_3}, \braket{g | e_4}\bigr]^\top$.
As a result, for the vector $\vec{s}$ of Eq.~(\ref{s_vec}), we have
$t_0^*\vec{t} = \braket{g | e_1}^*\vec{e} \sqrt{(1+3\alpha)(1-\alpha)}/4$ and
$\vec{t}^{\;*}\!\times \vec{t}=\vec{e}^{\;*}\!\times \vec{e}\, (1-\alpha)/4$. 
As expressed by Eqs.~(\ref{tr_ro})--(\ref{Qa1}), the vector $\vec{s}$, via the factor $Q_\xi$,
carries the coupling of the control qubit to the unitary $\mathsf{U}_\xi$, and therefrom the
capability of the control qubit for phase estimation on $\mathsf{U}_\xi$.
The important property here is that $\vec{s}$ vanishes, generically, with vanishing 
depolarizing noise at $\alpha =1$ where $\vec{t}=\vec{0}$, as discussed in 
Section~\ref{noiscoup_sec}; but on the contrary $\vec{s}$ does not vanish, generically, at 
maximum depolarizing noise when $\alpha =0$ where $\vec{t}\not =\vec{0}$.
In this way, with a fully depolarizing noise at $\alpha =0$, the capability for phase 
estimation of the controlled qubit of the superposed channel is generically preserved, for any 
environment model implementing the depolarizing noise.

\bigbreak 
For further comparison, it can be considered that the superposed channel, with a control qubit 
and a probe qubit, is a two-qubit process, even when measuring only the control qubit for phase 
estimation. Accordingly, a comparison is meaningful also with two-qubit conventional schemes for 
estimation \cite{Giovannetti06,Giovannetti11,Demkowicz14}. With two independent probing qubits, 
the performance of conventional schemes assessed by the Fisher information is additive, and is 
easily obtained from the characterization we performed above via 
Eqs.~(\ref{Fq_stand1})--(\ref{Fc_stand1}). With two entangled probing qubits, the performance of 
conventional schemes assessed by the Fisher information can be super-additive, benefiting from 
Heisenberg enhancement, which is nevertheless very fragile in the presence of noise 
\cite{Giovannetti11,Demkowicz14,Chapeau18}. But in any case, in the presence of a fully 
depolarizing noise $\mathcal{N}(\cdot)$, a standard interaction of the noisy unitary of
Fig.~\ref{figUxiN} with a probing qubit, always outputs the completely depolarized state 
$\mathrm{I}_2/2$ where any information about the unitary $\mathsf{U}_\xi$ is suppressed. 
Therefore, with two or even more independent or entangled probing qubits, conventional 
techniques will remain inoperative for phase estimation in the presence of fully depolarizing 
noise, while the control qubit of the superposed channel remains operative, as we have shown 
above.

\bigbreak
In the coherently superposed channel, after measurement of the control qubit, the probe qubit 
gets placed in the conditional state $\rho_\pm^{\rm post}$ of Eqs.~(\ref{ro_post})--(\ref{r_post}), 
which generally depends on the phase $\xi$, as expected for a qubit directly interacting with 
the unitary $\mathsf{U}_\xi$. Measuring $\rho_\pm^{\rm post}$ can therefore provide additional 
information to estimate $\xi$.
The performance upon measuring $\rho_\pm^{\rm post}$ is however more complicated to analyze and 
optimize, depending on the geometry as in Fig.~\ref{figRepz}, and the optimality conditions 
usually depend on which of the two conditional states $\rho_+^{\rm post}$ or $\rho_-^{\rm post}$ 
is obtained. An interesting property is that, with a fully depolarizing noise, the conditional 
Bloch vector $\vec{r}_\pm^{\rm \,post}$ of Eq.~(\ref{r_post}) remains dependent on the phase 
$\xi$, and can therefore be exploited for its estimation, although it is to more difficult to 
do it optimally.

Finally, if one resorts to measuring the probe qubit as well as the control qubit, a joint 
measurement of the two-qubit state $\mathcal{S}(\rho \otimes \rho_c)$ of Eq.~(\ref{Sgenqb}) 
could be envisaged as an alternative for estimation. This is a fortiori a more complicated
strategy, to analyze, to optimize and to implement. Interesting properties could nevertheless 
result, based on the capabilities revealed by the analysis of the control qubit of the 
superposed channel we performed here.

\section{Conclusion}

We have analyzed the coherently superposed channel of Fig.~\ref{figSwi1} when superposing two 
copies of a qubit unitary operator $\mathsf{U}_\xi$ affected by a general qubit noise as in
Fig.~\ref{figUxiN}.
A characterization has been developed for the joint state $\mathcal{S}(\rho \otimes \rho_c)$ in 
Eq.~(\ref{Sgenqb}) of the probe-control qubit pair of the coherently superposed channel.
As an application, an analysis of the superposed channel and its performance has been performed 
for the fundamental metrological task of phase estimation on the unitary $\mathsf{U}_\xi$ in the 
presence of noise. A comparison has also been made with conventional techniques of estimation 
and with the switched quantum channel with indefinite causal order studied in \cite{Chapeau21}
for a similar task of phase estimation. In particular, the present study complements the recent
study of \cite{Chapeau21}, so as to obtain a consistent analysis of the capabilities of 
coherently controlled channels, superposed or switched, for the fundamental metrological task 
of quantum phase estimation on a noisy unitary operator.

A first important observation here is that the control qubit of the superposed channel, although 
it never directly interacts with the unitary $\mathsf{U}_\xi$, can nevertheless be measured 
alone for effective estimation on $\mathsf{U}_\xi$, while discarding the probe qubit that 
interacts with the unitary $\mathsf{U}_\xi$. This property also occurs in the switched channel 
of \cite{Chapeau21}, but it never occurs with conventional techniques where auxiliary inactive 
qubits must be measured with the active probing qubits to be of some use for estimation. This 
property of a $\mathsf{U}_\xi$-dependent control qubit in the superposed channel, is generically 
established in broad conditions here, for any qubit noise $\mathcal{N}(\cdot)$ in 
Fig.~\ref{figUxiN} and underlying environment model. In addition, the optimal measurement of the 
control qubit with maximum efficiency for phase estimation, has been characterized in general 
conditions on the noise and environment.

A second important observation is that the noise plays an essential role in coupling the
control qubit to the unitary $\mathsf{U}_\xi$, with usually a nonzero amount of noise
maximizing the efficiency of the control qubit for phase estimation on $\mathsf{U}_\xi$.
This is also established to hold generically in the superposed channel, for any qubit noise and 
environment model. In addition, it is observed that the control qubit of the superposed channel 
remains operative for phase estimation at very strong noise, even with a fully depolarizing 
noise with any implementing environment, whereas conventional estimation techniques and the 
switched channel with indefinite order of \cite{Chapeau21} become inoperative in these conditions. 
This represents here a specificity of the superposed channel that is significant for noisy quantum
metrology. The fully depolarizing noise, 
which does not suppress the estimation capability in the superposed channel here, is 
reminiscent of the completely depolarizing channel that enables information communication when 
placed in coherently controlled associations, switched or superposed, as shown in 
\cite{Ebler18,Loizeau20,Abbott20}. However, the tasks and performance metrics are different for 
estimation and communication, and the present results provide additional elements for broader 
appreciation of the capabilities and specificities of coherently controlled channels 
--- switched as well as superposed --- as novel devices exploitable for various tasks of 
quantum information processing.

The characterization of the joint probe-control state $\mathcal{S}(\rho \otimes \rho_c)$ 
carried out in Section~\ref{qbswitch_sec} could be further exploited for other purposes
involving the noisy unitary placed in a coherent superposition. Estimation of the axis 
$\vec{n}$ of the unitary could as well be envisaged, and due to the fundamental coupling 
described by Eqs.~(\ref{tr_ro})--(\ref{Qa1}) via the $\mathsf{U}_\xi$-dependent factor 
$Q_\xi$, such estimation again would be possible by measuring the control qubit alone.

As we have seen in the analysis, the operation of the coherently superposed channel and the joint 
probe-control state $\mathcal{S}(\rho \otimes \rho_c)$ of Eq.~(\ref{Sgen2}) it outputs, depend
on the underlying implementation of the non-unitary channels realized by the environment. 
When superposing two identical qubit channels, as done for phase estimation on $\mathsf{U}_\xi$,
this dependence can be encapsulated in four real scalar parameters 
$(s_0, \vec{s}\,)$, independently of the size and constitutive details of the environment. If 
these parameters $(s_0, \vec{s}\,)$ cannot be deduced from prior knowledge or
assumptions available for the underlying environment, they can be included in the 
estimation process analyzed in the study and can be estimated from measurement of the control 
qubit of the superposed channel. Such an estimation process of a few scalar parameters for the 
environment in fact provides a handle enabling to put in practical use coherently superposed
qubit channels, for phase estimation as addressed here, but also for information communication 
as addressed in \cite{Abbott20,Loizeau20}, or for other tasks of quantum signal and information 
processing. For instance, in practice, based on the present analysis, a unitary process 
invincibly hidden in a fully depolarizing noise, could become amenable to estimation via its 
duplication and insertion in a interferometric setup as in \cite{Zhou11,Friis14,Abbott20} 
experimentally realizing the controlled coherent superposition. Other possibilities are opened 
for exploration by exploiting coherently controlled channels for quantum signal and information 
processing.

\section*{Appendix}

\renewcommand{\theequation}{A-\arabic{equation}} 
\setcounter{equation}{0}  

In this Appendix we provide additional elements on the Stinespring implementation of quantum 
channels, like channels (1) and (2) of Section~\ref{switch_sec}, and having special relevance 
for their coherent superposition as in Fig.~\ref{figSwi1}. On a quantum system $Q$ with Hilbert 
space $\mathcal{H}$, a quantum operation or channel $\rho \mapsto \mathcal{E}(\rho)$ defined by a 
set of $K$ Kraus operators $\{\mathsf{K}_j \}_{j=1}^{K}$ of $\mathcal{L}(\mathcal{H})$, can be 
obtained \cite{Nielsen00,Wilde17} from an environment model $E$ with a $K$-dimensional Hilbert 
space $\mathcal{H}_E$ implementing on the joint system-environment compound $QE$ the (dilated) 
unitary evolution
\begin{equation}
\ket{\psi}\otimes\ket{g} \xmapsto{\displaystyle \;\mathsf{U_J}\;} 
\mathsf{U_J} \ket{\psi}\otimes \ket{g}=\ket{\phi_{QE}} \in \mathcal{H} \otimes \mathcal{H}_E\;,
\label{A_U1}
\end{equation}  
for any state $\ket{\psi} \in \mathcal{H}$ of system $Q$ and where 
$\ket{g} \in \mathcal{H}_E$ is the initial state of the environment $E$. An orthonormal basis 
$\{\ket{e_j}\}_{j=1}^{K}$ of the $K$-dimensional environment space $\mathcal{H}_E$ is used for 
performing partial tracing over the environment, in order to obtain the quantum operation on 
system $Q$ as $\rho =\ket{\psi}\bra{\psi} \mapsto \mathcal{E}(\rho)
=\tr_{E}\bigl( \ket{\phi_{QE}}\bra{\phi_{QE}} \bigr)=
\sum_{j=1}^K \braket{e_j | \phi_{QE}} \braket{\phi_{QE}| e_j}$. From Eq.~(\ref{A_U1}) we obtain 
in this way the operator-sum representation of the quantum operation as
$\mathcal{E}(\rho)= \sum_{j=1}^K  \mathsf{K}_j \ket{\psi}\bra{\psi}\mathsf{K}_j^\dagger$,
defining the Kraus operators
\begin{equation}
\mathsf{K}_j=\big\langle e_j| \mathsf{U_J} | g \big\rangle \;,
\label{A_Kraus1}
\end{equation}  
for $j=1$ to $K$, via the notion of partial inner product \cite{Schumacher10,Nielsen00}. By 
linearity, the same operator-sum representation of $\mathcal{E}(\rho)$ applies when $\rho$ is a 
mixed state as a convex sum of pure states like $\ket{\psi}\bra{\psi}$. Then Eq.~(\ref{A_Kraus1}) 
enables us to write an equivalent form for the joint unitary evolution $\mathsf{U_J}$ of the 
system-environment compound $QE$ of Eq.~(\ref{A_U1}) as
\begin{equation}
\ket{\psi}\otimes\ket{g} \xmapsto{\displaystyle \;\mathsf{U_J}\;} 
\sum_{j=1}^K \mathsf{K}_j \ket{\psi} \otimes \ket{e_j} =
\ket{\phi_{QE}} \in \mathcal{H} \otimes \mathcal{H}_E \;,
\label{A_U2}
\end{equation}  
forming the ground for the evolutions of Eqs.~(\ref{environ1})--(\ref{environ2}). For a given
quantum operation $\mathcal{E}(\cdot)$, the minimal number of Kraus operators is determined by 
the rank of $\mathcal{E}(\cdot)$ and is no larger than $[\text{dim}(\mathcal{H})]^2$.
Any two sets $\{\mathsf{K}_j \}_{j=1}^{K}$ and $\{\mathsf{K}'_{k} \}_{k=1}^{K'}$ of Kraus 
operators offer an equivalent operator-sum representation of the same quantum operation 
$\mathcal{E}(\cdot)$, if and only if they can be connected \cite{Nielsen00} via a unitary 
transformation 
\begin{equation}
\mathsf{K}'_{k} = \sum_{j=1}^{K'} u_{k j} \mathsf{K}_j\;,
\label{2.Kequiv1}
\end{equation}
where the smaller set of the two has been complemented with null operators to equalize the sizes 
$K$ and $K'$, and the complex numbers $[u_{k j}]$ form a (square) unitary matrix, in particular 
satisfying $\sum_\ell u_{\ell j} u^*_{\ell k} =\delta_{jk}$ and 
$\sum_\ell u_{j \ell} u^*_{k \ell} =\delta_{jk}$, so that Eq.~(\ref{2.Kequiv1}) inverts as
$\mathsf{K}_j = \sum_k u^*_{k j} \mathsf{K}'_{k}$.

A given environment model fixed by $\bigl(\ket{g}, \mathsf{U_J}\bigr)$ in Eq.~(\ref{A_U1}),
with dimension $K$, generates the set $\{\mathsf{K}_{j} \}_{j=1}^{K}$ of $K$ Kraus 
operators via Eq.~(\ref{A_Kraus1}) when the $K$-dimensional environment space $\mathcal{H}_E$
is referred to the orthonormal basis $\{\ket{e_j} \}_{j=1}^{K}$. From here, a larger 
set $\{\mathsf{K}'_{k} \}_{k=1}^{K'}$ of $K' \ge K$ Kraus operators offering an equivalent 
representation for the same quantum operation $\mathcal{E}(\cdot)$, can be obtained via 
Eq.~(\ref{2.Kequiv1}), for any size $K' \ge K$, by fixing a unitary matrix $[u_{k j}]$. This 
larger set $\{\mathsf{K}'_{k} \}_{k=1}^{K'}$ of $K' \ge K$ Kraus operators can be associated 
with an environment model $E'$, which, according to the logic above, will be tied to a 
larger Hilbert space $\mathcal{H}'_E$ with dimension $K' \ge K$. This Hilbert space 
$\mathcal{H}'_E$ can be selected in such a way that $\mathcal{H}_E$ with dimension $K$ for the 
environment $E$, is a subspace of $\mathcal{H}'_E$ with dimension $K' \ge K$ for the environment 
$E'$, this corresponding to enlarging the number of degrees of freedom of the environment coupled 
to system $Q$.

From the unitary connection of Eq.~(\ref{2.Kequiv1}), the set
$\{\mathsf{K}'_{k} \}_{k=1}^{K'}$ is recovered by the partial products
$\mathsf{K}'_{k}=\big\langle e'_{k}| \mathsf{U'_J} | g \big\rangle$ 
by referring to the orthonormal basis $\{\ket{e'_{k}} \}_{k=1}^{K'}$ of $\mathcal{H}'_E$.
This basis $\{\ket{e'_{k}} \}_{k=1}^{K'}$ of $\mathcal{H}'_E$ is connected to the
basis $\{\ket{e_{j}} \}_{j=1}^{K}$ of $\mathcal{H}_E$ via the unitary
transformation
\begin{equation}
\ket{e'_{k}} = \sum_{j=1}^{K'} u^*_{k j} \ket{e_j} \Longleftrightarrow
\ket{e_{j}} = \sum_{k=1}^{K'} u_{k j} \ket{e'_{k}} \;,
\label{2.basis2b}
\end{equation}
where the smaller basis $\{\ket{e_{j}} \}_{j=1}^{K}$ of 
$\mathcal{H}_E \subseteq \mathcal{H}'_E$ has been complemented (in a non-critical way) by $K'-K$ 
orthonormal states $\{\ket{e_{j}} \}_{j=K+1}^{K'}$ to reach an orthonormal basis of
$\mathcal{H}'_E$. In a comparable way, the unitary $\mathsf{U'_J}$ acting in the larger space
$\mathcal{H} \otimes \mathcal{H}'_E$, has an action coinciding with $\mathsf{U_J}$ in 
$\mathcal{H} \otimes \mathcal{H}_E$, and is complemented outside in a non-critical way. The 
initial state $\ket{g} \in \mathcal{H}_E \subseteq \mathcal{H}'_E$ does not change while 
enlarging the environment dimensionality, and the evolution of the system-environment compound 
of Eq.~(\ref{A_U2}) can be written in the two alternative forms
\begin{equation}
\ket{\psi}\otimes\ket{g} \longmapsto
\sum_{j=1}^K \mathsf{K}_j \ket{\psi} \otimes \ket{e_j} =
\sum_{k=1}^{K'} \mathsf{K}'_{k} \ket{\psi} \otimes \ket{e'_k} =\ket{\phi_{QE}} 
\in \mathcal{H} \otimes \mathcal{H}_E \subseteq \mathcal{H} \otimes \mathcal{H}'_E \;.
\label{A_U3}
\end{equation}  

In this way, starting from a set $\{\mathsf{K}_{j} \}_{j=1}^{K}$ of $K$ Kraus operators and its 
generative environment model, we can deduce, by introducing an arbitrary unitary matrix
$[u_{k j}]$, an infinite number of equivalent Kraus sets $\{\mathsf{K}'_{k} \}_{k=1}^{K'}$ along 
with their generative environment model, and representing the same quantum operation 
$\mathcal{E}(\cdot)$. The significant feature for the superposed channel of Fig.~\ref{figSwi1} 
is that all these equivalent representations share the same transformation operators like 
$\mathsf{T}_1$ and $\mathsf{T}_2$ of Eqs.~(\ref{opT1})--(\ref{opT2}), since by the
unitary connections of Eqs.~(\ref{2.Kequiv1})--(\ref{2.basis2b}) we generically have
\begin{equation}
\mathsf{T}_1 = \sum_{j=1}^{K} \braket{g | e_j} \mathsf{K}_j 
=\sum_{k=1}^{K'} \braket{g | e'_{k}} \mathsf{K'}_{k} \;.
\label{opTu}
\end{equation}
Moreover, by expressing in Eq.~(\ref{opTu}) the Kraus operators by their partial product 
expression, we finally obtain $\mathsf{T}_1 =\big\langle g| \mathsf{U_J} | g \big\rangle$,
manifesting the invariance of the transformation operators of Eqs.~(\ref{opT1})--(\ref{opT2}) in
the change of Kraus operators. This is the same as 
$\mathsf{T}_1 =\big\langle g| \mathsf{U'_J} | g \big\rangle =
\big\langle g| \mathsf{U_J} | g \big\rangle$ for any enlarged unitary $\mathsf{U'_J}$ since 
their action on $\ket{g}$ all coincide. In this respect, the specific choice of one set of 
Kraus operators with associated environment model, among all equivalent sets for channels (1) 
and (2) as described above, will not change $\mathsf{T}_1$ and $\mathsf{T}_2$ in
Eqs.~(\ref{opT1})--(\ref{opT2}), and therefore will not affect the operation of the superposed 
channel of Fig.~\ref{figSwi1}. The modeling problem is to have access, for each channel (1) and 
(2), at least to one convenient pair of Kraus set and environment model, among the infinite 
number of equivalent pairs, so as to have access to a characterization of the transformation 
operators $\mathsf{T}_1$ and $\mathsf{T}_2$ of Eqs.~(\ref{opT1})--(\ref{opT2}) governing the 
coherent superposition in Fig.~\ref{figSwi1}. This may require, for each channel (1) and (2), 
beyond a set of Kraus operators, additional knowledge or assumptions concerning the underlying 
implementation. Alternatively, in the absence of such constitutive informations or modeling 
assumptions, an estimation procedure can be envisaged of the unknown channel parameters, as 
examined in Section~\ref{implem_sec}. Besides, in the paper, for qubit channels, we work out a 
general description having common validity for arbitrary channel parameters.


\end{document}